\documentclass[twocolumn]{emulateapj}
\usepackage{xcolor}

\usepackage{lineno}

\shorttitle{Understanding the role of $\omega$ Cen building up the Milky Way halo}
\shortauthors{Anguiano et al.}

\begin{document}

\title{Tidal Debris Candidates from the $\omega$ Centauri Accretion Event and its Role in Building Up the Milky Way Halo\\[-1.0cm]}

\author{Borja Anguiano$^{\dagger,\ddagger}$\altaffilmark{1,2}, Arik Mitschang\altaffilmark{3}, Takanobu Kirihara\altaffilmark{4}, Yutaka Hirai\altaffilmark{5}, Danny Horta\altaffilmark{6}, Sten Hasselquist\altaffilmark{7}, Ricardo P. Schiavon\altaffilmark{8}, Steven R. Majewski\altaffilmark{2}, Andrew C. Mason\altaffilmark{8}, Adrian M. Price-Whelan\altaffilmark{6}, Carlos Allende Prieto\altaffilmark{9}, Verne Smith\altaffilmark{10}, Katia Cunha\altaffilmark{11}, David L. Nidever\altaffilmark{12}}

\altaffiltext{1}{Centro de Estudios de F\'isica del Cosmos de Arag\'on (CEFCA), Plaza San Juan 1, 44001, Teruel, Spain}
\altaffiltext{2}{Department of Astronomy, University of Virginia, Charlottesville, VA, 22904, USA}
\altaffiltext{3}{Henry A. Rowland Department of Physics and Astronomy, The Johns Hopkins University, Baltimore, MD 21218, USA}
\altaffiltext{4}{Kitami Institute of Technology, 165, Koen-cho, Kitami, Hokkaido 090-8507, Japan}
\altaffiltext{5}{Department of Community Service and Science, Tohoku University of Community Service and Science, 3-5-1 Iimoriyama, Sakata, Yamagata 998-8580, Japan}
\altaffiltext{6}{Center for Computational Astrophysics, Flatiron Institute, 162 Fifth Avenue, New York, NY 10010, USA}
\altaffiltext{7}{Space Telescope Science Institute, 3700 San Martin Drive, Baltimore, MD 21218, USA}
\altaffiltext{8}{Astrophysics Research Institute, Liverpool John Moores University, 146 Brownlow Hill, Liverpool L3 5RF, UK}
\altaffiltext{9}{Instituto de Astrofísica de Canarias, E-38205 La Laguna, Tenerife, Spain; Universidad de La Laguna, Dpto. Astrofísica, E-38206 La Laguna, Tenerife, Spain}
\altaffiltext{10}{National Optical Astronomy Observatories, Tucson, AZ 85719, USA}
\altaffiltext{11}{Steward Observatory, University of Arizona, 933 North Cherry Avenue, Tucson, AZ 85721-0065, USA}
\altaffiltext{12}{Department of Physics, Montana State University, PO Box 173840, Bozeman, MT 59717-3840, USA}

\thanks{Ram\'on y Cajal Fellow}
\email{banguiano@cefca.es}

\begin{abstract}
We identify stellar tidal debris from the $\omega$ Centauri ($\omega$ Cen) system among field stars in the APOGEE survey via chemical tagging using a neural network trained on APOGEE observations of the $\omega$ Cen core. We find a total of 463 $\omega$ Cen debris candidates have a probability $P > 0.8$ of sharing 
common patterns in their chemical abundances across a range of individual elements or element combinations, including [C+N], O, Mg, Al, Si, Ca, Ni, and Fe. Some debris candidates show prograde or retrograde disk-like kinematics, but most show kinematics consistent with the accreted halo, showing high radial actions, $J_{R}$, values. We find that a sample of Gaia-Sausage-Enceladus (GES) members are chemically distinct from the $\omega$ Cen core, suggesting that $\omega$ Cen is associated to an independent merger event shaping the Milky Way halo. However, a connection between GSE and $\omega$ Cen cannot be ruled out. A detailed comparison with $N$-body simulations indicates that the $\omega$ Cen progenitor was a massive dwarf galaxy ($\gtrsim 10^8 M_{\odot}$). The existence of a metal-poor high-$\alpha$ chemically homogeneous halo debris is also reported. 

\end{abstract}

    
\section{Introduction} 
\label{sec:intro}

The ``globular cluster" $\omega$ Centauri ($\omega$ Cen) has a number of peculiar characteristics that set it apart from other Milky Way globular clusters; these include its very large mass, extended size, oblate shape, internal rotation, large age and metallicity spreads \citep{Piotto2005,JP2010,Bellini2010,Villanova2014}, and a retrograde orbit only slightly inclined to the Galactic plane \citep{Dinescu2002,Vasiliev2019}. Because of these properties, together with its peculiar surface brightness profile, it is believed that $\omega$ Cen may be the core remnant of a heavily stripped, Milky Way-captured dwarf spheroidal galaxy, now currently orbiting retrograde near the Galactic plane \citep[e.g.,][]{Lee1999,Majewski2000,Bekki2003,Tsuchiya2003,Tsuchiya2004,Ideta2004}. In a few, generally massive globular clusters where $\omega$ Cen represents the most extreme case, different internal stellar populations show abundance variations of virtually all elements, suggesting that these massive clusters might be considered as an intermediate evolutionary stage between globular clusters and ultra compact dwarf galaxies \citep[e.g.,][]{Forbes&Kroupa2011,Norris&Kannappan2011}. Additionally, \cite{Noyola2008} detected a rise in velocity dispersion from 18.6 km s$^{-1}$ at 14 arcsec to 23 km s$^{-1}$ in the center, a rise associated with the existence of an intermediate mass black hole. \cite{Cheng2020}, using X-ray sources as sensitive probes of stellar dynamical interactions, also reported the existence of a black hole subsystem in $\omega$ Cen. Recently, \cite{Haberle2024}, using fast-moving stars in the central region of the cluster, reported the existence of an intermediate-mass black hole with a mass of $\sim$ 8,200 M$_{\odot}$. Moreover, potentially accreted stellar systems have been reported within the cluster \citep[e.g.,][]{Ferraro2002,Calamida2020}. These findings imply that $\omega$ Cen has galaxy-like properties. 

There is now strong evidence to indicate active tidal disruption of $\omega$ Cen, with the detection of tidal tails extending directly from the system finally identified by \cite{Ibata2019,Kuzma2025}, after years of negative, uncertain or low-significance detections for this expected phenomenon \citep[e.g.,][]{Leon2000,Law2003,DaCosta2008,Wylie2010,Anguiano2015,FT2015}. \cite{Ibata2019} also reported that the Fimbulthul structure is a tidal stream of $\omega$ Cen (see also \citealt{Simpson2020}).
Furthermore, the main chemical properties of $\omega$ Cen can be reproduced if it is the compact remnant of a dwarf spheroidal galaxy that evolved in isolation and then was accreted and partly disrupted by the Milky Way's tidal forces. 
The ingested satellite would also have chemical properties similar to the core of $\omega$ Cen \citep{Romano2007}.  $N$-body/gas dynamical simulations of the formation and evolution of $\omega$ Cen \citep{Carraro2000,Tsuchiya2004} reproduce the bulk properties of the cluster --- namely its structure, kinematics and chemistry --- assuming that this stellar system formed and evolved in isolation, and eventually fell inside the Milky Way potential well. \cite{Carraro2000} concluded that $\omega$ Cen could actually be a cosmological dwarf by mass, a system formed in a high redshift low mass halo that escaped significant merging up to the present time. Moreover, \cite{Meza2005} used numerical simulations to analyze the dynamical properties of tidal debris stripped from a satellite on a highly eccentric orbit, and they found that these satellites may deposit a significant fraction of their stars in the disk components of a galaxy. Furthermore, using a compilations of nearby metal-poor stars, \cite{Meza2005} reported a stellar group that based on dynamical and chemical coherence, appears to consist of stars that once belonged to the dwarf that brought $\omega$ Cen into the Galaxy.
 
In the context of the connection between globular clusters and dwarf galaxies, it is widely accepted that the globular cluster M54 is associated with the Sagittarius dwarf spheroidal galaxy (Sgr) and that M54
formed and evolved in the environment of a dwarf spheroidal galaxy \citep[e.g.,][]{Sarajedini1995,Majewski2003,Monaco2005,Carretta2010,Bellazzini2008}. 
The M54/Sgr system is often suggested as the paradigm for a putative former $\omega$ Cen progenitor, and this has inspired the search for tidal debris shed from said progenitor now lying among Milky Way field stars.  For example, \citet{Dinescu2002} looked for metal-poor stars in the solar neighborhood that shared $\omega$ Cen's orbit.
More recently, \cite{Myeong2018}, using Gaia DR2 \citep{GaiaDR2},  discovered a new retrograde substructure in the Milky Way halo (named the ``Sequoia" galaxy), 
that they associated with the $\omega$ Cen progenitor. Their analysis shows the efficacy of using kinematics to identify potential $\omega$ Cen stars.  
However, as we shall show in \S\ref{sec:data}, retrograde behavior alone --- and even orbits of a common angular momentum --- are not sufficient evidence of a linkage to $\omega$ Cen.  For this reason, additional characteristics --- such as chemistry --- are needed to discriminate $\omega$ Cen debris.
Alternatively, \citet{Myeong2018} and \citet{Massari2019} studied the age-metallicity relation of Milky Way globular clusters associated with the different halo structures and concluded that $\omega$ Cen could have been the nuclear star cluster of the more massive Gaia-Sausage-Enceladus galaxy. 

Previous searches for more extended $\omega$ Cen tidal debris within a large, all-sky low-resolution spectroscopic and photometric catalog of giant stars by \citet{Majewski12} identified candidate retrograde stars near the Galactic plane having kinematics (derived solely from radial velocities) consistent with being stripped debris from $\omega$ Cen, based on tidal destruction models of the system. To confirm their status as former $\omega$ Cen members, high resolution spectroscopy was undertaken of a dozen of these candidates, and most were found by \citet{Majewski12} to exhibit very high relative Ba abundances (as measured by the $\lambda$5854 \AA\  transition) --- a peculiar characteristic apparently unique to the $\omega$ Cen system, as originally shown by \cite{Norris&DaCosta1995} and \citet{Smith2000}. See also Figure 11 in \cite{Geisler2007}. Thus, these results showed the likelihood of a connection between a set of widely distributed field stars and $\omega$ Cen, which could then be used to make some broad inferences about the orbit of the progenitor.

Although generating a relatively small, though broadly outlying group of $\omega$ Cen debris stars, the \citet{Majewski12} study demonstrated an effective procedure for selecting candidates according to their chemical fingerprint. Such procedure can be useful for creating a larger census of $\omega$ Cen stars across the Milky Way.
With the advent of large-scale Milky Way stellar surveys like {\it Gaia} and APOGEE providing plentiful and accurate chemo-kinematical data for vastly large numbers of stars, a far more comprehensive search for $\omega$ Cen debris stars can be mounted following a similar prescription. In this study, we performed such a larger analysis using accurate multi-element chemical abundances from the Apache Point Observatory Galactic Evolution Experiment (APOGEE, \citealt{Majewski2017}) to search for tidal debris candidates from the $\omega$ Cen accretion event. 

The concept of chemical tagging consists in using detailed chemical abundances of individual stars to tag them to common ancient star-forming aggregates whose stars have similar abundance patterns \citep[e.g.,][]{Freeman&Bland-Hawthorn2002,Mitschang2013,Mitschang2014,DeSilva2015}. The feasibility of chemical tagging has been investigated with promising results \citep[e.g.,][]{Quillen2015, Hogg2016, Kos2018,Spina2022}. However, \cite{Garcia_Dias2019,Casamiquela21} showed that different stellar birth sites can have overlapping chemical signatures, suggesting substantial evidence against the feasibility of a strong chemical tagging. Recently, \cite{Youakim2023} used the combined GALAH DR3  \citep{GALAHDR3} and \emph{Gaia} eDR3 \citep{GaiaEDR3} data sets to identify candidates stripped off from $\omega$ Cen. Using simultaneously the chemical and dynamical parameter space via an unsupervised clustering algorithm (t-SNE, \citealt{tsne}), \cite{Youakim2023} reported debris candidates extending more than 50$^{\circ}$ away from the cluster. Recently, \cite{Pagnini2025} proposed that at least six Milky Way globular clusters share a common origin with $\omega$ Cen, further supporting the hypothesis that this system is the remnant of a galaxy.

In a similar vein, from the combination of full 3-D kinematics derived from {\it Gaia} proper motions and APOGEE radial velocities, accurate multi-element chemical abundances from APOGEE, and distances from the StarHorse value-added catalog \citep{SH2020}, we identify a large sample of several hundred stars sharing $\omega$ Cen's peculiar chemical characteristics and use them to constrain the present and past orbital properties of the $\omega$ Cen progenitor. With our new sample of $\omega$ Cen stars thus identify, a more complete and definitive mapping of $\omega$ Cen tidal debris can be built and used to shed further light on its complex chemo-dynamical evolution. In particular,
we conjecture about the possible association of $\omega$ Cen with other, previously identified,  major accreted structures in the Milky Way halo. 

\section{APOGEE and Gaia datasets}
\label{sec:data}

The Apache Point Observatory Galactic Evolution Experiment (APOGEE, \citealt{Majewski2017}), part of both SDSS-III \citep{Eisenstein2011} and SDSS-IV \citep{Blanton2017}, explored the multi-element chemistry and radial velocities of stars across all stellar populations of the Milky Way using high-resolution ($R \sim 22,500$) spectra spanning a large fraction of the $H$-band (1.5-1.7$\mu$m).  APOGEE in SDSS-III, initially operating from the Sloan 2.5 m Telescope \citep{Gunn2006} at Apache Point Observatory, expanded observations to include both hemispheres as part of APOGEE-2 in SDSS-IV through the installation of a twin spectrograph \citep{Wilson2019} on the du Pont telescope at Las Campanas Observatory. In addition, the new APOGEE-South spectrograph was used to procure observations of almost two thousand
red giant branch (RGB) stars in the $\omega$ Cen core (including observations of the system during ``first light'' commissiong of the instrument --- e.g., \citealt{Meszaros2021}).  These spectra of the $\omega$ Cen core provide critical templates for our Galaxy-wide search for stripped $\omega$ Cen counterparts.  

\begin{figure}[ht]
\begin{center}
\includegraphics[width=1.12\hsize,angle=0]{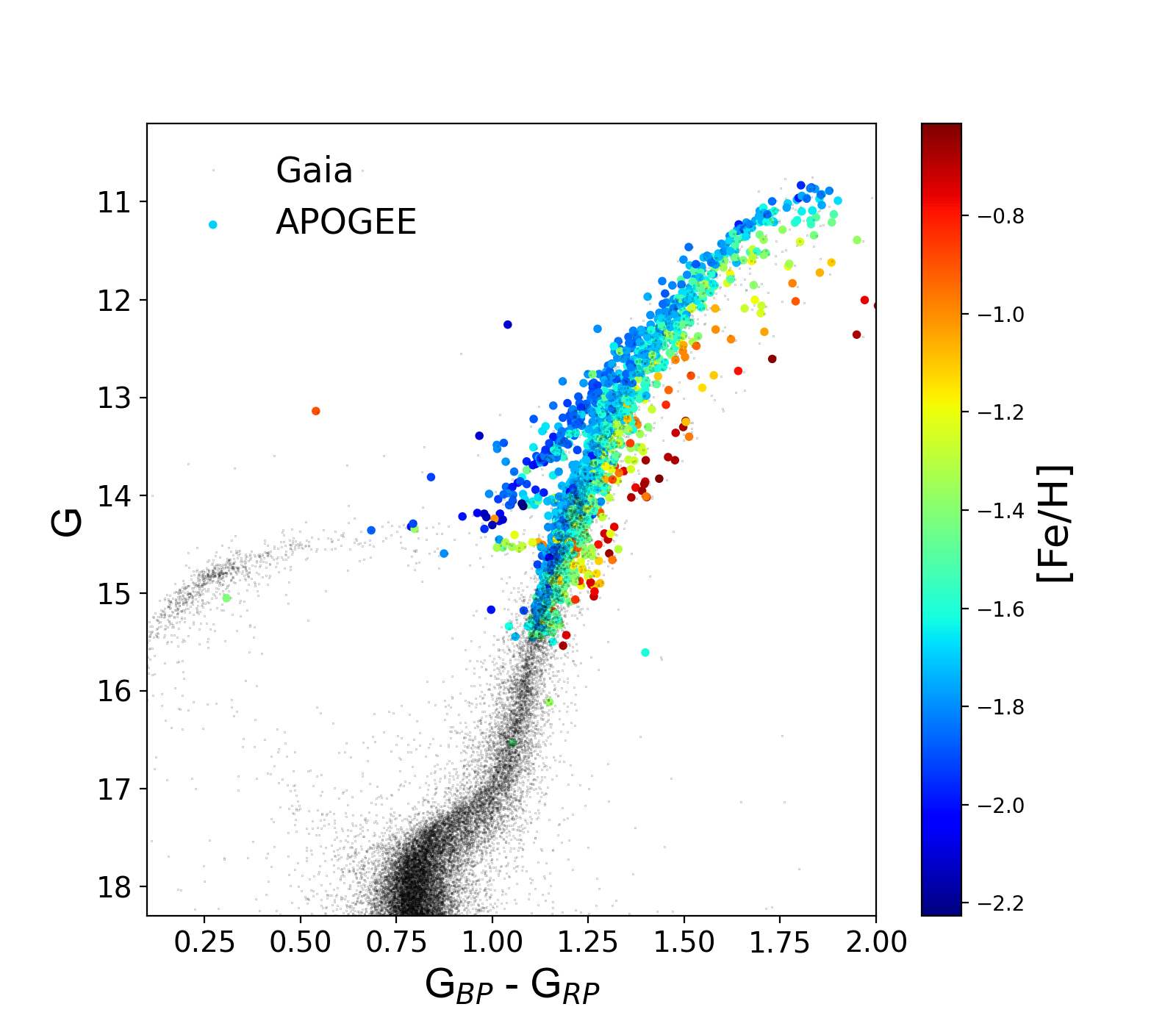}
\end{center}
\caption{{\it Gaia}-based color-magnitude diagram of the $\omega$ Cen system, with the APOGEE targets marked, color-coded by the ASPCAP-derived [Fe/H]. APOGEE observations are covering the wide
distribution of the $\omega$ Cen red-giant branch stars and also the wide metallicity spread.}
\label{fig_CMD}
\end{figure}

We make use of the last SDSS-IV public release of data from the APOGEE Stellar Parameters and Chemical Abundances Pipeline (ASPCAP, \citealt{Garcia_Perez2016, Holtzman2018}) based on MARCS model atmospheres \citep{Gustafsson2008} , \verb|SYNSPEC| \citep{Hubeny2021}, and \verb|FERRE| \citep{FERRE2015}. These data were processed and publicly released within SDSS Data Release 17 (DR17, \citealt{SDSS_DR17})\footnote{\url{https://www.sdss4.org/dr17/irspec/spectro\_data/}}. Using the APOGEE line list \citep{Shetrone2015, Hasselquist2016, Cunha2017, Smith2021}, the DR17 reduction pipelines deliver $v_{\rm los}$, stellar atmospheric parameters, and abundances for 18 individual chemical species \citep{Nidever2015,Holtzman2015,Garcia_Perez2016,Jonsson2020}, some of them considering departures from LTE \citep{Osorio2020}.

In this study, we also exploit data from the third data release (DR3) of the \emph{Gaia} mission \citep{Gaia2022} through the APOGEE StarHorse value-added catalog \citep{SH2020}. We use the Galactocentric frame in Astropy v5.2.1 \citep{Astropy2022}, which contains the following parameters: a solar Galactocentric distance of $R_{0}$ = 8.275 kpc, a solar motion in cylindrical coordinates of ($\upsilon_{R,\odot}$, $\upsilon_{\phi,\odot}$,$\upsilon_{z,\odot}$) = (8.4, 251.8, 8.4) km s$^{-1}$, and a solar position with respect to the Galactic midplane of $Z_{\odot}$ = 20.8 pc. We use v1.3 of gala\footnote{https://gala.adrian.pw/en/latest/} \citep{PW2017} to compute actions and other orbital information, and for which we adopt \emph{Gaia} DR3 astrometry, StarHorse distances, and APOGEE DR17 radial velocities. We used gala's default gravitational potential from \cite{Bovy2015galpy}, which is matched to the \cite{Eilers2019} rotation curve. In this way, we put together a catalog of orbital properties, such as eccentricities, peri/apocenter radii, maximal disk height $Z$, orbital actions, frequencies, and angles, together with their uncertainties for all stars. The adopted reference frame is right-handed, i.e., stars with $L_{z}$ $<$ 0 are in prograde orbits, while those with $L_{z}$ $>$ 0 are retrograde. 

\begin{figure}[ht]
\begin{center}
\includegraphics[width=1.0\hsize,angle=0]{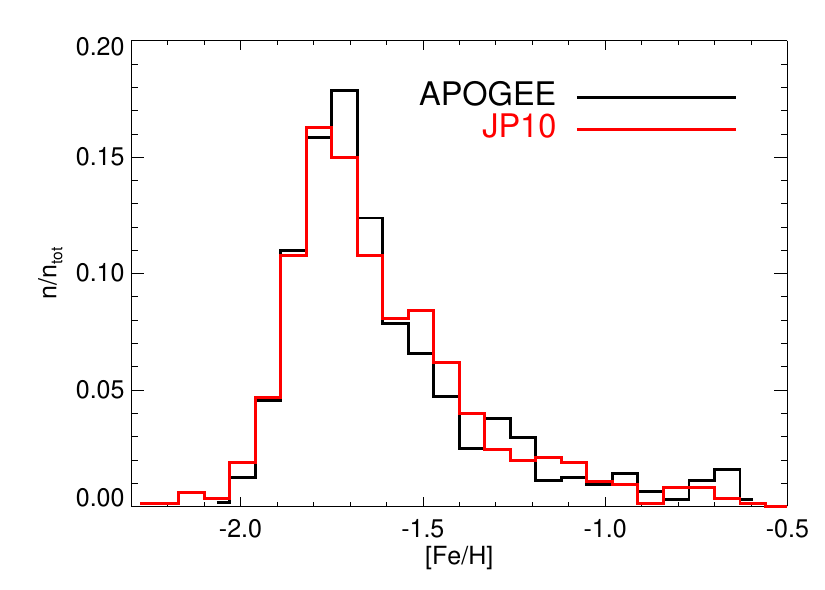}
\includegraphics[width=.95\hsize,angle=0]{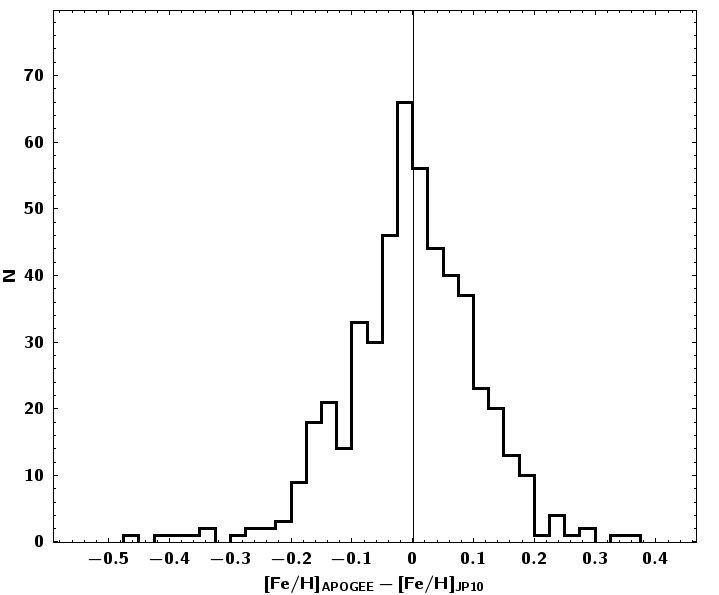}
\end{center}
\caption{\emph{Top panel:} Normalized fraction of the metallicity distribution function of $\omega$ Cen populations for the APOGEE (black) and \citet[][``JP10'']{JP2010} (red) sample. There is an agreement between the APOGEE and JP10 [Fe/H] distributions. \emph{Bottom panel:} Histogram of the discrepancies between the [Fe/H] measurements from APOGEE spectra and the values from JP10. The two independent measurements show good agreement, with a standard deviation of just 0.1.}
\label{fig:fe_ocen}
\end{figure}

\subsection{APOGEE Observations of the $\omega$ Centauri Core}
\label{core}

If our goal is to find stellar debris far flung from the $\omega$ Cen core, we need to rely on stellar properties that are not only characteristic of the system but also conserved. Stellar chemical abundance patterns are a property characteristic of parent systems that (for the most part) is maintained by their constituent stars for most of their lives.  Fortunately, the $\omega$ Cen core has received special attention in APOGEE-2 \citep{Santana2021}, and these observations can be used to look for APOGEE-based chemical patterns that define $\omega$ Cen.


Stars in the $\omega$ Cen core were observed as faint as $H$=13.5. As is routine for APOGEE, the nominal minimum S/N is 100; nevertheless, for the $\omega$ Cen data analyzed here, including commissioning data obtained before the instrument and observing procedures were fully optimized, the range of S/N spans 40 to 300. A total of 1,872 $\omega$ Cen stars had APOGEE spectra that were successfully processed through the APOGEE Stellar Parameters and Chemical Abundances Pipeline \citep[ASPCAP;][]{Garcia_Perez2016}. For further information about the $\omega$ Cen targeting strategy we refer the reader to \cite{Santana2021}.

\begin{figure*}[ht]
\begin{center}
\includegraphics[width=1.\hsize,angle=0]{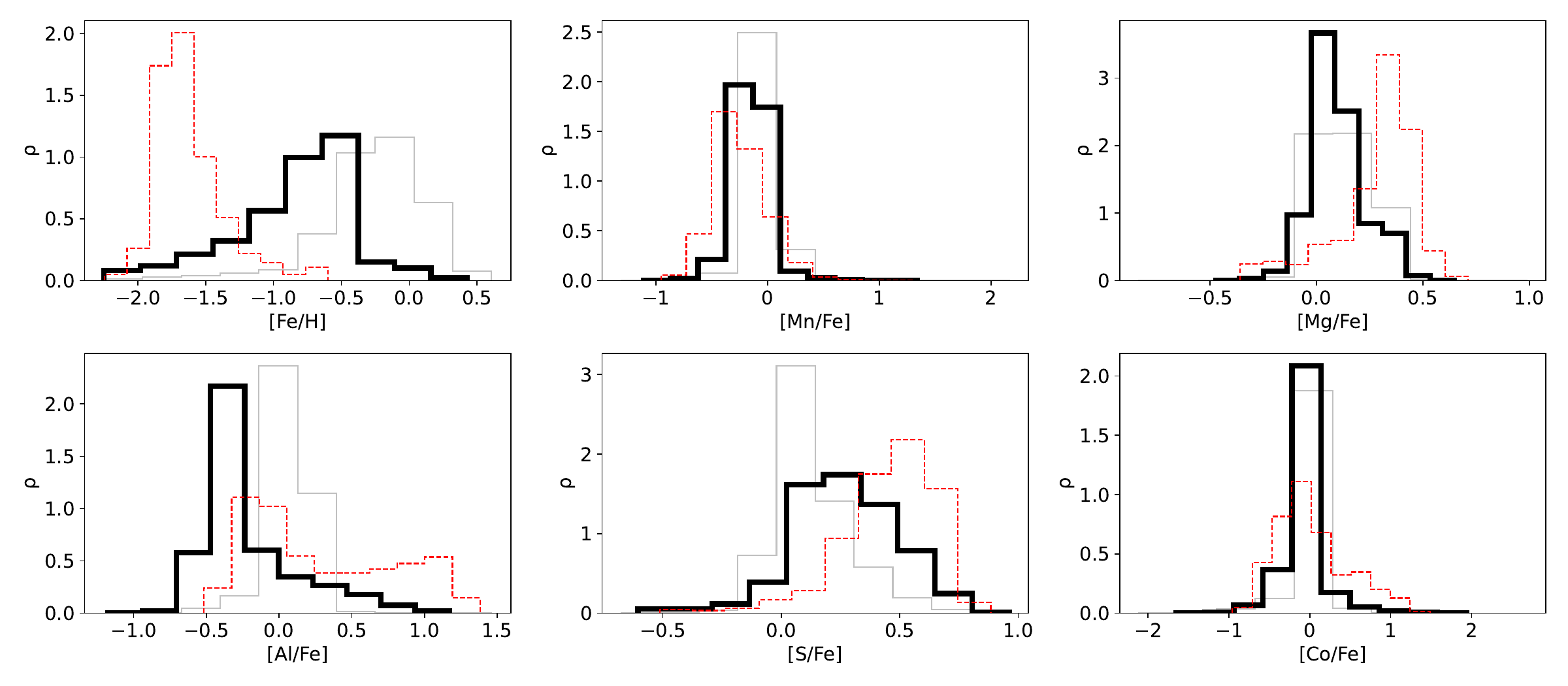}
\end{center}
\caption{Distributions of the chemical tagging model negative training sample (solid black line), compared to the $\omega$ Cen subsample that represent the positive training data (dashed red line) and the full APOGEE sample (grey line) for six individual abundances.}
\label{fig_input_hist_vs_abund}
\end{figure*}

Figure~\ref{fig_CMD} shows a {\it Gaia}-based CMD of the system, with the APOGEE targets marked, color-coded by the ASPCAP-derived [Fe/H].  As may be seen by the wide distribution of the $\omega$ Cen RGB stars in the CMD and the APOGEE-based metallicity distribution function (MDF) of these stars shown in Figure~\ref{fig:fe_ocen} , the APOGEE targets span the full range of $\omega$ Cen populations ($-2.3$ $<$ [Fe/H] $<$ $-0.5$).  This wide coverage ensures that we sample the diverse range of $\omega$ Cen chemistries. Those chemistries allow us to search for abundance signatures of $\omega$ Cen directly as measured by APOGEE. Some of the unusual chemical properties exhibited by $\omega$ Cen have been highlighted in \citet{Meszaros20,Meszaros2021}.

In Figure~\ref{fig:fe_ocen} we also show the [Fe/H] distribution of \citet[][``JP10'']{JP2010}. There is excellent agreement between the APOGEE and JP10 [Fe/H] distributions, with 511 stars in common between the two samples. The figure also presents the histogram of discrepancies between the two independent measurements. The comparison indicates a strong agreement, with a minimal offset of 0.01 and a standard deviation of just 0.1. Although every GC studied so far harbors at least two stellar generations distinct in chemical composition among the light elements \citep{Gratton2004} and also age, GCs are still found to be mono-metallic objects as far as abundances of heavier elements are concerned.\footnote{This is apart from a few notable confirmed exceptions: M54 \citep{Bellazzini2008}, M22 \citep{Marino2009}, and M2 \citep{Yong2014}. See Table 10 in \cite{Marino2015}, for a summary of clusters with iron spreads.} The most notable exception is $\omega$ Cen, where evidence of several bursts of star formations with corresponding peaks in the metal abundance are reported \citep[e.g.,][]{Norris&DaCosta1995,JP2010,Gratton2011,Meszaros2021}. In the next section, we discuss how we use the chemical peculiarities of $\omega$ Cen to identify debris candidates in the APOGEE survey. 


\begin{figure}[ht]
\begin{center}
\includegraphics[width=1.\hsize,angle=0]{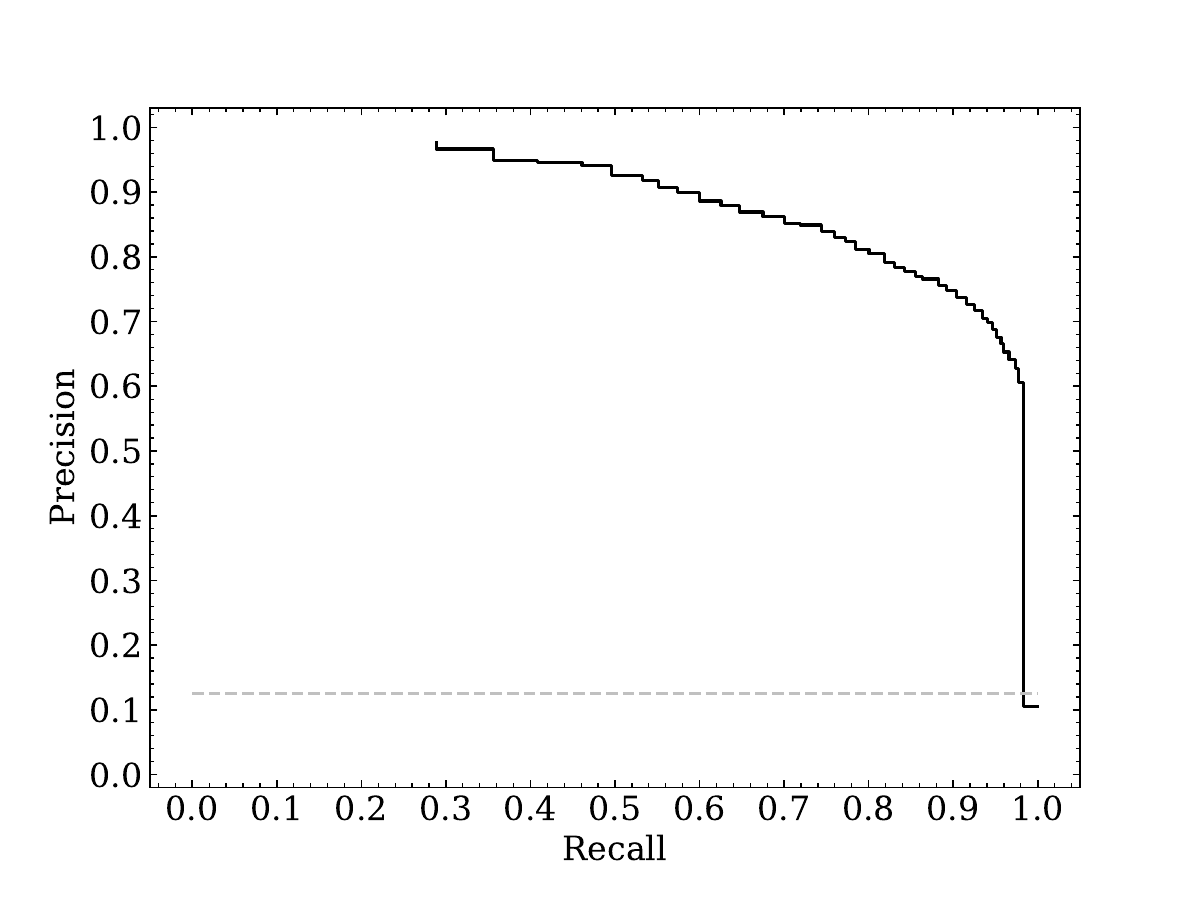}
\end{center}
\caption{Neural network model performance. The best model recall and precision are plotted for $P > P_{thresh}$ from 0 to 1 at intervals of 0.01. The dashed grey horizontal line at ~0.12 represents a random uneducated model. Our best fit does substantially better, achieving close to 80\% recall at 80\% precision.}
\label{fig_recall_precision}
\end{figure}

\section{$\omega$ Cen debris candidates}
\label{debris_candi}

As discussed, GCs are known to display very uniform metallicities and prominent light element abundance correlations and anti-correlations. The most well-known of these are the Na-O, Mg-Al and C-N anti-correlations detected in most of the GCs analyzed so far \citep[e.g.,][]{Carretta2009b,Carretta2009,Pancino2017}. In the case of $\omega$ Cen, its unusual chemistry requires an evolution different from that followed by any known globular cluster. Previous studies show the complexity of the abundance pattern, together with its multiple populations \citep{Denissenkov1998,Pancino2000,JP2010,Gratton2011,Marino2012,Meszaros2021}, suggesting that $\omega$ Cen was massive enough to retain the ejecta of supernovae and the subsequent enrichment of new stellar generations with iron produced at the end of each burst of star formation. In principle, these distinct chemical signatures make GC stars stripped off their original cluster easily identifiable \citep[e.g.,][and references therein]{Martell2016,Schiavon2017,Horta2021,Sian2022}. 

\cite{Freeman&Bland-Hawthorn2002} discussed the concept of \emph{chemical tagging}, in which stars are linked to individual star formation events when their abundance patterns in a range of elements, from $\alpha$ to Fe-peak, light to heavy $s$- and $r$-process, are the same \citep[e.g.,][]{Mitschang2014}. Chemical tagging techniques have already been explored using the APOGEE data set \citep[e.g.,][]{Hogg2016,Ness2017,Andrews2019,Webb2020}, and also the so-called \emph{weak} chemical tagging, where we chemically label stars born in the same type of stellar system rather than
the same unique stellar system \citep[e.g.,][]{Schiavon2017,Fernandez-Trincado2017,Hayes2018a,Hasselquist2019}.

In this section, we explain how we select debris candidates using \emph{weak} chemical tagging from a neural network model that has the APOGEE $\omega$ Cen core as a training sample. We also explore the chemo-dynamical properties of the debris candidates. 

\subsection{Chemical Selection of Debris Candidates in the APOGEE Survey: Chemical Tagging Model}
\label{sec:chemtagmodel}

As a first step to identifying $\omega$ Cen candidates, we search for stars that share common patterns in their chemical abundances across a range of elements, in hopes that these so-called ``chemical tags" are unique to the evolutionary origins of $\omega$ Cen (chemical tagging references). Instead of specifically modeling the numerous complex processes that underlie the chemical abundance patterns of $\omega$ Cen progenitors, we employ a black-box model capable of identifying arbitrary, potentially non-linear relationships across a large number of dimensions. For this we use a ``neural network" \citep[e.g.,][]{NN}. We built the training set using the $\omega$ Cen core observed by APOGEE (see Section~\ref{core} for details). We select the $\omega$ Cen stars for which the signal-to-noise ratio (S/N) exceeds 75 and no \verb|ASPCAPBAD| and \verb|STARFLAG| are set \citep{Holtzman2015}. This yields a total of 1794 objects for the training set. The same selection in terms of (S/N) and APOGEE flags is applied to the entire APOGEE sample. Figure~\ref{fig_input_hist_vs_abund} shows the distribution of six abundances used in the chemical tagging model for the training set (dashed red line), the model training data (solid black line), and the full APOGEE sample (grey line).  


For our search for $\omega$ Cen tidal debris, we do not include focused APOGEE stars in science programs that target specific objects such as the Sgr dSph galaxy \citep{Majewski2013,Hasselquist2019,Hayes2020}, the Large and Small Magellanic Clouds \citep{Nidever2020}, numerous star clusters different from $\omega$ Cen \citep{Meszaros2020,schiavon2024}, and the dwarf spheroidal galaxy programs \citep{Zasowski2017,Beaton2021,Santana2021} that are not germane to this study of field stars. To remove these specialized targets from our data base, we identified the fields associated with the special programs and removed all the targets within those fields from our search database.




We perform the chemical tagging of candidate $\omega$ Cen debris by training a neural network model to generate a probability-like indicator of membership in $\omega$ Cen. Those stars determined by the model to be chemically similar to $\omega$ Cen have a higher value of this indicator, which can then be used to fine-tune the false positive rate of candidates in tagging. The neural network model we construct is of the well-known fully connected multi-layer perceptron (MLP) type, having an input layer representing the abundance measurements, three hidden layers of neurons with 256, 128, and 32 units, respectively, and an output layer containing, in effect, a single unit representing the membership probability. Each edge is a transfer function $\omega x_{in} + \beta$, and each neuron applies an activation function $f(x)$ to its input and sums the results, which are then inputs to transfer functions of the next layer. For the two hidden layers, we use the \emph{Rectified Linear Unit} (RelU) activation function, and for the final layer, we apply the \emph{Softmax} function, which transforms the input to a probability of class membership, adding to 1 across the inputs. Training proceeds by computing the error on output for each training sample --- where the error function used is the mean squared error --- and adjusting the weights via a gradient descent technique so as to minimize the error.

To achieve a sufficient generalization of the model such that it could be used to tag unknown stars in our APOGEE sample, we need to train it with sufficient examples of both the target ($\omega$ Cen) and representative negative examples. We include 1454 stars from 27 known globular clusters present in our filtered APOGEE sample \citep{schiavon2024}, in addition to 5434 stars from the LMC and the SMC \citep{Nidever2020}, and 400 identified as members of the Sagittarius dSph or stream \citep{Hayes2020}. We note that while the globular cluster sample provides negative examples, there is also an expectation that these are chemically homogeneous to a higher degree than ``field" stars and thus are more separable, leading to an overly optimistic view of performance if treated alone. The LMC, SMC and Sgr stars provide a more well-mixed background with more complex enrichment histories built-in. These combined form the so-called negative sample that we use to train the chemical tagging model. In total we have 15119 objects that comprise the negative sample.

The model is operating entirely in  chemical space. Following the discussion described in \cite{Holtzman2018} and \cite{ SDSS_DR17}, we use the following individual abundances measured in the APOGEE spectra for our exercise: [C/Fe], [N/Fe], [O/Fe], [Mg/Fe], [Al/Fe], [Si/Fe], [Mn/Fe], [Ni/Fe], [Fe/H], [CI/Fe], [Na/Fe], [K/Fe], [Ca/Fe], [Co/Fe], [Ce/Fe]. Of these, Mg, Si, Na, K and Ca are derived in NLTE by the ASPCAP pipeline. To avoid introducing significant bias through the imputation of missing abundance measurements, we excluded any element for which the fraction of missing values in the negative sample differed by more than 25 percentage points from that in the $\omega$ Cen sample. This exercise removed [Na/Fe] and [Ce/Fe] from the candidate abundances. Table \ref{model_abundance_sample_inputs} shows the final list of abundances used and coverage among the target and negative samples.

\begin{deluxetable*}{lrrrrrrrrrrrrrr}
\centering
\tablewidth{0pt}
\tabletypesize{\footnotesize}
\tablecaption{Model input abundances coverage \label{model_abundance_sample_inputs}}
\tablecolumns{15}
\tablehead{
\colhead{Sample} & \colhead{[Si/Fe]} & \colhead{[Fe/H]} & 
\colhead{[Mg/Fe]} & \colhead{[Ni/Fe]} & \colhead{[Al/Fe]} & \colhead{[O/Fe]} & \colhead{[Ca/Fe]} & \colhead{[CI/Fe]} & \colhead{[K/Fe]} & \colhead{[C/Fe]} & \colhead{[CO/Fe]} & \colhead{[N/Fe]} & \colhead{[Mn/Fe]}
}
\startdata
$\omega$ Cen    &   1.00 &  1.00 &   0.99 &   0.99 &   0.99 &  0.97 &   0.94 &   0.90 &  0.89 &  0.88 &   0.81 &  0.69 &   0.67 \\
Negative &   1.00 &  0.99 &   1.00 &   0.99 &   0.87 &  0.99 &   0.99 &   0.97 &  0.96 &  0.97 &   0.95 &  0.89 &   0.63 \\

\enddata
\end{deluxetable*}


Figure \ref{fig_recall_precision} shows the performance of the model in terms of recall (the proportion of cluster stars that were correctly identified as such, the true positive rate) and precision (the proportion of positive labels that were correctly marked positive) as a function of threshold in probability ($P > P_{thresh}$).
We use cross-validation. A random model -- one that makes an uneducated guess for each input -- would have a constant precision equal to the proportion of positive samples in the dataset for all thresholds, while recall would increase linearly with decreasing threshold, achieving 1 and a threshold of 0 (simply assuming everything is a member).


\begin{figure}[ht]
\begin{center}
\includegraphics[width=1.\hsize,angle=0]{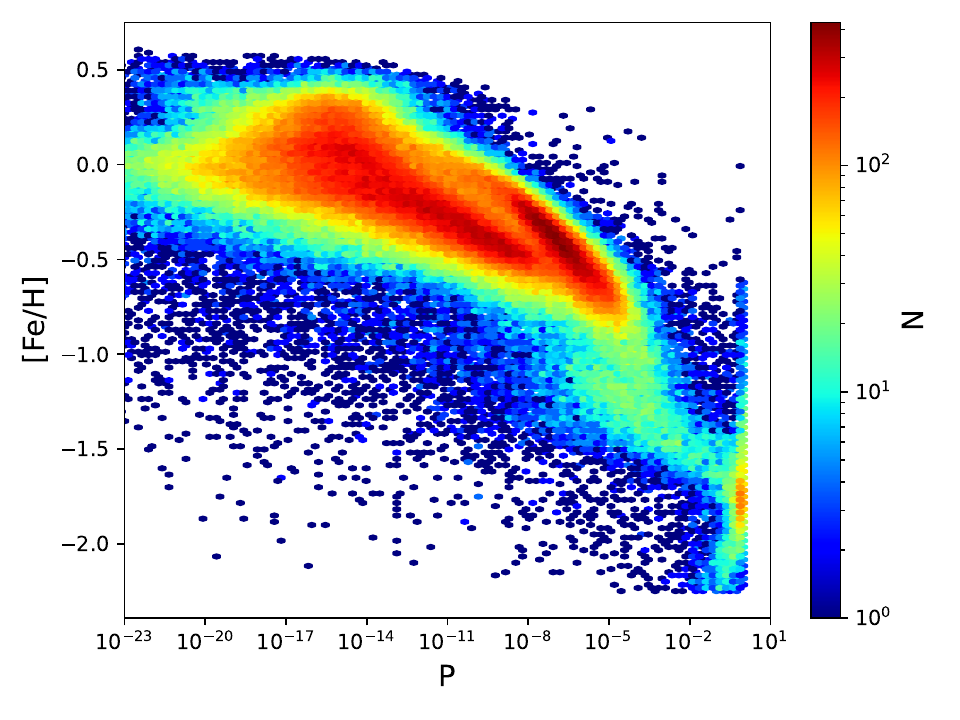}
\end{center}
\caption{[Fe/H] as a function of chemically tagged membership probability. Note the built-in dependence on [Fe/H] abundance and the resulting MW structure detected by the model. Also note the heavy tendency for low probabilities -- more than 99\% of stars in the sample have $P < 0.01$.}
\label{fig_log_prob_vs_feh}
\end{figure}

\begin{figure*}[ht]
\begin{center}
\includegraphics[width=.7\hsize,angle=0]{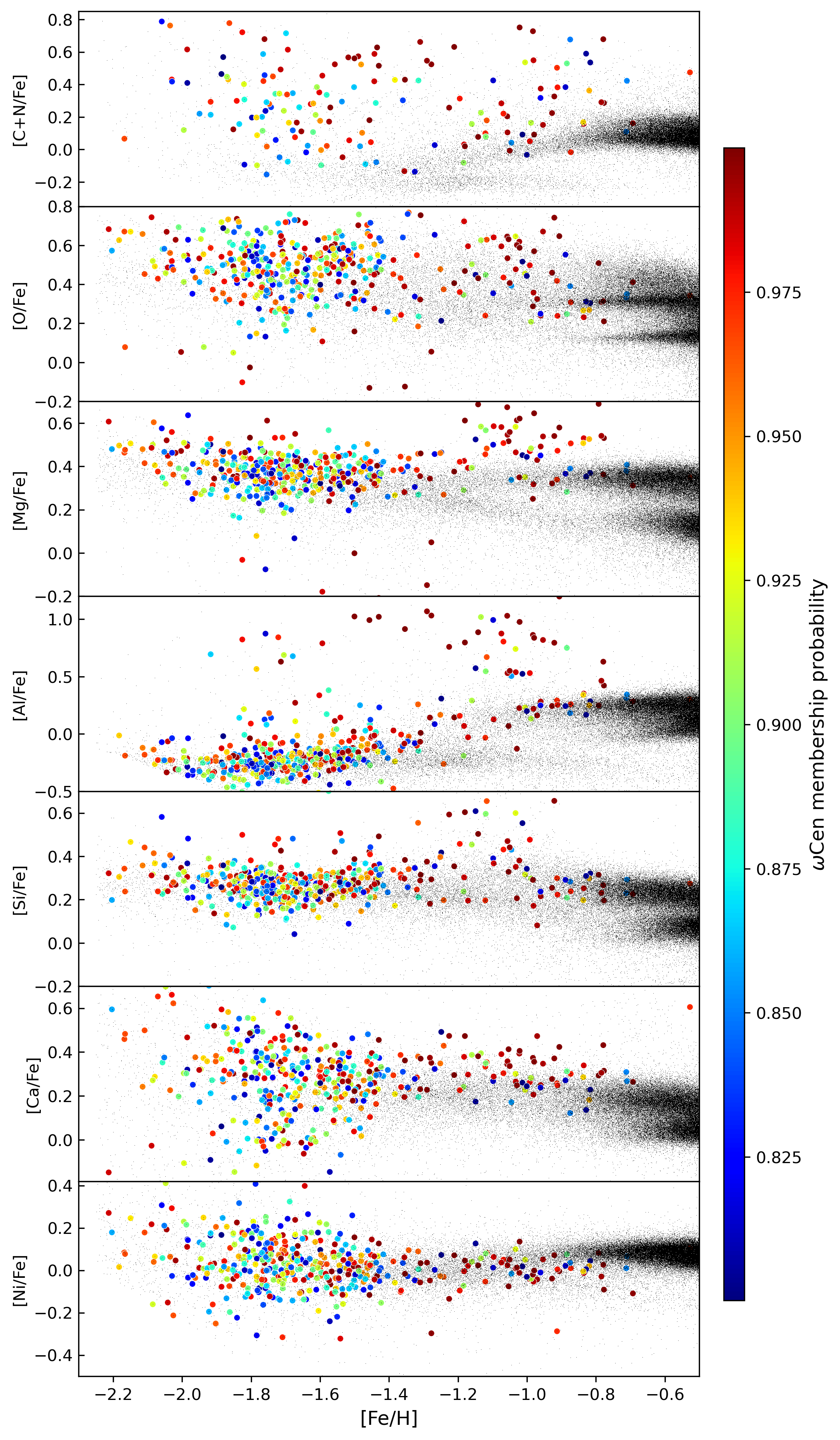}
\end{center}
\caption{Chemical abundance patterns of select elements for $\omega$ Cen debris candidates as compared to the MW. The debris candidates are color-coded by the $\omega$ Cen membership probabilities, and we only show candidates where $P > 0.8$}
\label{fig_ELE_P}
\end{figure*}

\begin{figure*}[ht]
\begin{center}
\includegraphics[width=1.\hsize,angle=0]{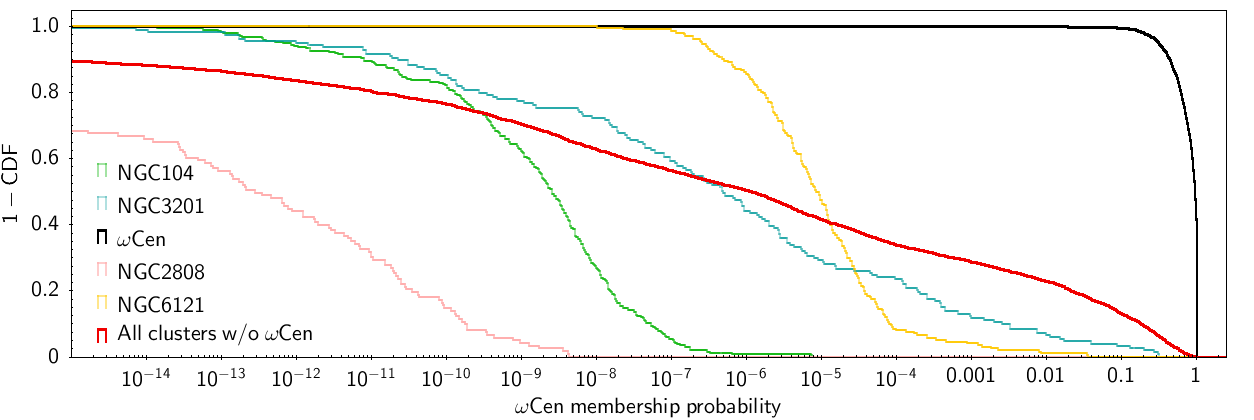}
\end{center}
\caption{Inverse cumulative distribution functions for the $\omega$ Cen membership probability for four distinct clusters (NGC104, NGC3201, NGC2808, and NGC6121), as well as $\omega$ Cen itself (black line). We also present the CDF for every member of the globular cluster listed in APOGEE DR17, excluding the $\omega$ Cen members. Although 90$\%$ of the members of other groups have an extremely low probability of being part of $\omega$ Cen (red line), according to our chemical selection, approximately 10$\%$ of the total of members of other clusters could have a probability of 0.1, and only 1$\%$ a probability of 0.6 to be part of $\omega$ Cen.}
\label{CDF}
\end{figure*}

\begin{figure*}[ht]
\begin{center}
\includegraphics[width=.48\hsize,angle=0]{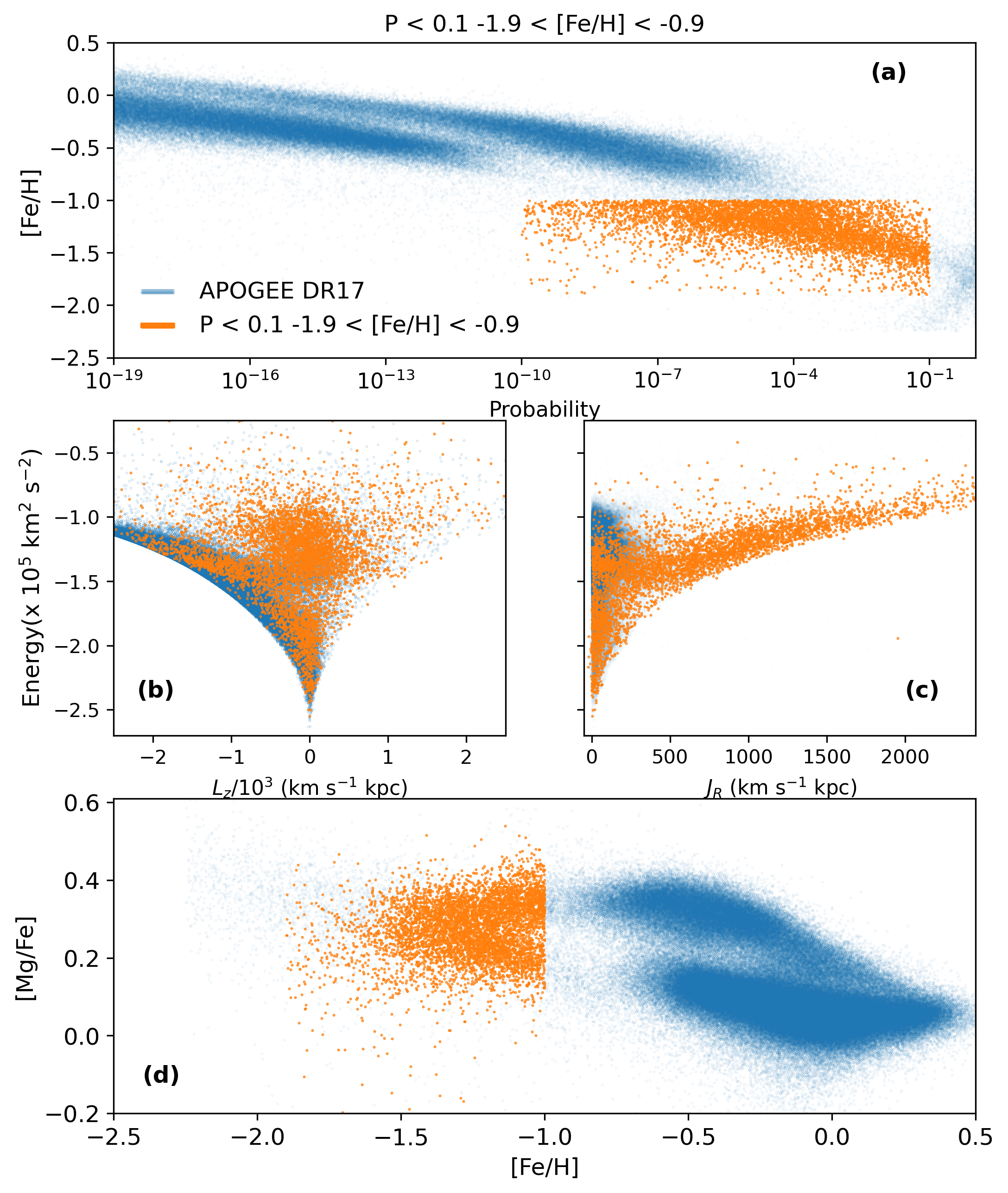}
\includegraphics[width=.48\hsize,angle=0]{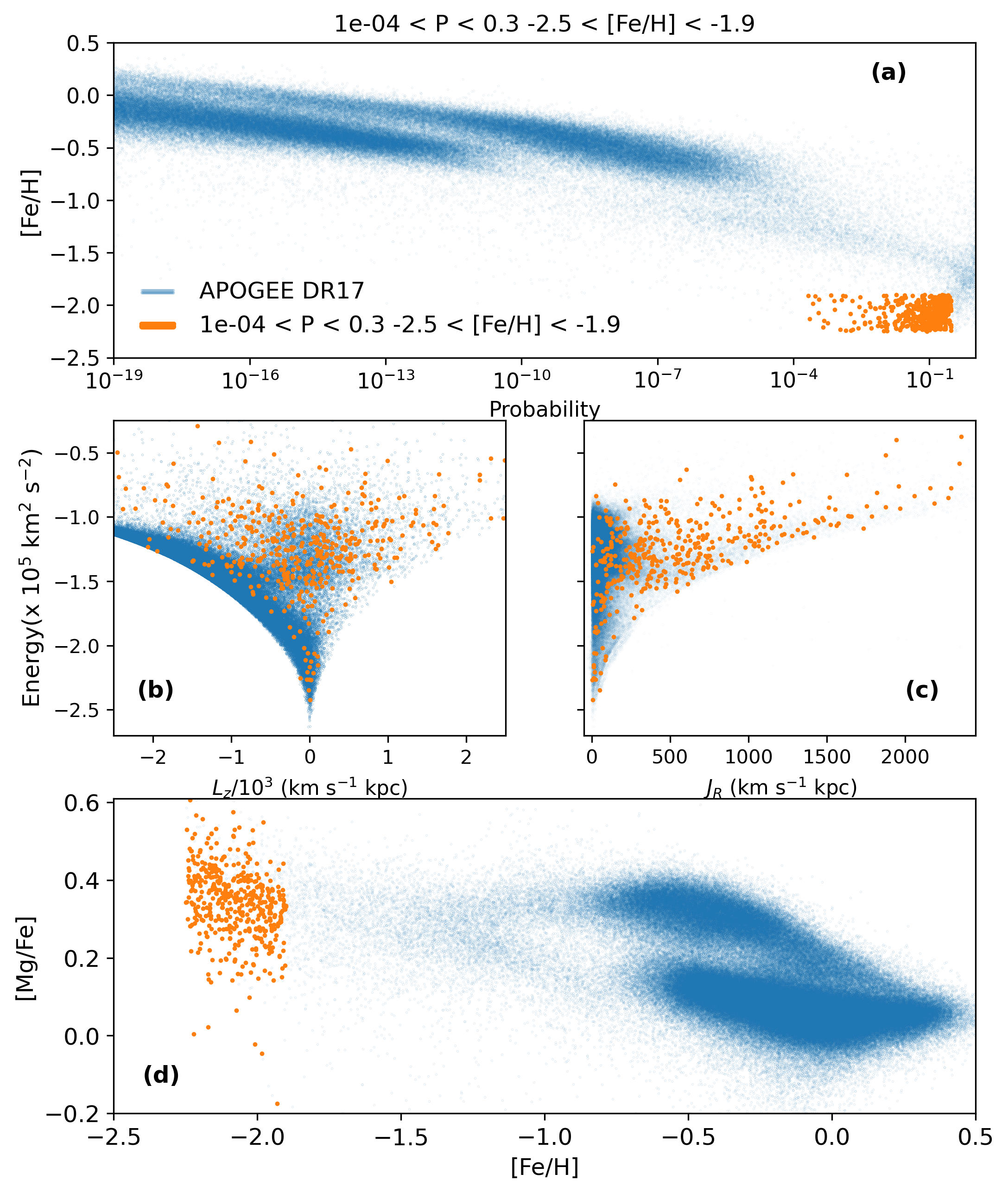}
\end{center}
\caption{$\omega$ Cen membership probability as a function of [Fe/H], energy-action spaces and the [Fe/H] - [Mg/Fe] plane. Left-hand panels: the orange color in these panels highlights the selection of $P < 0.1$ and $-$1.9 $<$ [Fe/H] $<$ $-$ 0.9. There are clearly three structures, a metal-weak thick disk, a non-rotating low orbital energy population, and the halo debris associated with GSE. We discuss them in detail in the text. Right-hand panels: the orange color represent the selection for the metal-poor ``plume", that is very likely an ancient merger in the MW halo. In all the panels, the blue data represent the APOGEE DR17 catalog. The figures helps to visualize the chemo-dynamical study of the different structures identify in this study.} 
\label{fig_P_limb}
\end{figure*}

\begin{figure}[ht]
\begin{center}
\includegraphics[width=1.\hsize,angle=0]{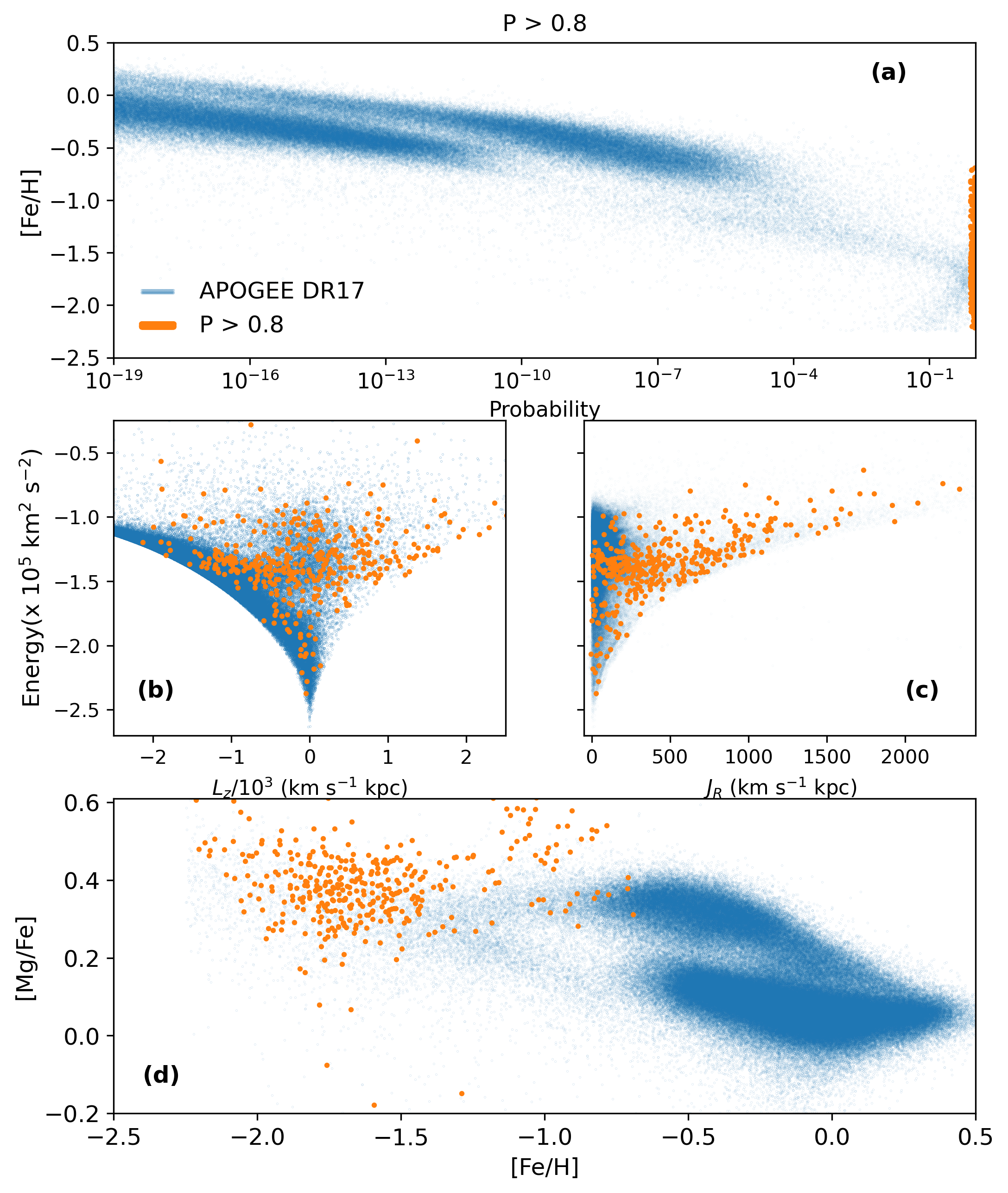}
\end{center}
\caption{$\omega$ Cen membership probability as a function of [Fe/H], energy-action spaces and the [Fe/H] - [Mg/Fe] plane for the $\omega$ Cen debris candidates with $P > 0.8$. The Lindblad diagram reveals that most of the debris lie within the accreted halo with some having disk-like kinematics.}
\label{fig_P_ocen}
\end{figure}


\begin{deluxetable*}{lcc}
\centering
\tablewidth{0pt}
\tabletypesize{\footnotesize}
\tablecaption{Column Descriptions for the $\omega$ Cen debris candidates \label{debris_candi}}
\tablecolumns{3}
\tablehead{
\colhead{Column Name} & \colhead{Data Type} & \colhead{Description} \\
}
\startdata
target\_id & String & Unique source identifier \\
apogee\_id & String & APOGEE identifier for the star \\
C\_FE & Double & Carbon-to-Iron abundance ratio \\
N\_FE & Double & Nitrogen-to-Iron abundance ratio \\
O\_FE & Double & Oxygen-to-Iron abundance ratio \\
... & ... & ... \\
pred\_ocen & Double & $\omega$ Cen membership probability \\
\enddata
\tablecomments{This table lists the metadata for the selected columns from the oCen\_debris\_APOGEE\_Anguiano.fits file. It provides column names, data types, and brief descriptions for each parameter. The full data set is available in the electronic edition of the {\it Astrophysical Journal}.}
\end{deluxetable*}


\subsection{Chemo-dynamical Study of Debris Candidates}
\label{sec:data}

Figure~\ref{fig_log_prob_vs_feh} shows the stellar APOGEE [Fe/H] as a function of $\omega$ Cen chemically tagged membership probability. In this figure, we clearly detect the structure related to the MW disk and halo at very low probabilities. We also have a prominent over-density close to $P$ $\sim$ 1 and ranging from $-$2.1 $<$ [Fe/H] $<$ $-$1.2. Interestingly, there is a very metal-poor ``plume" ([Fe/H] $<$ $-$2.0 ) with a $P$ $\sim$ 0.1 following the chemical space of $\omega$ Cen, but distinct from the remaining substructures identified in the figure. 

We now explore the chemical abundance patterns for C+N,\footnote{We combine C and N because stars change their [C/Fe] and [N/Fe] abundances during dredge-up ascending to and along the giant branch, but these processes occur in such a way that the [(C+N)/Fe] abundance remains largely constant before and after these mixing processes \citep[e.g.,][]{Gratton2000}.}
the $\alpha$-elements (O, Mg, Si, Ca), Al, Ni, and Ce for the $\omega$ Cen debris candidates (see Figure~\ref{fig_ELE_P}). For all of these exercises, the $\omega$ Cen core observed by APOGEE (see Sect.~\ref{core}) is removed. For a detailed chemical abundance analysis $\omega$ Cen core stars using APOGEE abundances, we refer the reader to \cite{Meszaros2021} and Mason et al. (2025, submitted). We select these elements because they are among the most precise APOGEE abundances across the full parameter space covered here, and are among the most accurate when comparing to optical studies \citep[e.g.,][]{Jonsson2020,Hasselquist2021}. 

The data points representing $\omega$~Cen debris candidates in Figure~\ref{fig_ELE_P} are color-coded by the $\omega$ Cen membership probability, where we show the debris candidates with $P$ $>$ 0.8. We find that the number of $\omega$ Cen debris candidates with $P$ $>$ 0.8 is 463, while for $P$ $>$ 0.9 we have 284 candidates and for $P$ $>$ 0.95 the total number is 186 objects. Table~\ref{debris_candi} shows the first four entries of the debris candidates with $P$ $>$ 0.8. 


The results in Figure~\ref{fig_ELE_P} show that there is an overlap, especially for O, Mg, Si, Ca, and Ni, between the multiple populations of the $\omega$ Cen debris candidates and the field stars in the Milky Way halo. 
This shows the efficiency of the neural network employed in this study (Sect.~\ref{sec:chemtagmodel}) to discern populations with different chemical evolution histories. 
A good number of the debris candidates are clearly enhanced in C+N, Al abundances with respect to the rest of the debris population and most of the Galactic halo. For instance, the high [Al/Fe] abundances we see in some globular cluster stars implies that a previous generation of higher-mass or evolved stars must have contributed to their chemical composition \citep{Charbonnel2006}. 
We also see this \emph{excess} in other light elements, like in the C+N panel. 
The hallmark C+N and Al trends are potential evidence that an extended part of the $\omega$ Cen tidal debris stream has been found. 

Furthermore, we try to understand the optimal selection for $\omega$ Cen membership probability. Figure~\ref{CDF} shows the reverse Cumulative Distribution Function (CDF) for the $\omega$ Cen 
membership probability 
of $\omega$ Cen (black line) and four clusters for which APOGEE observed more than 200 individual members. It is remarkable how chemically unique all clusters are compared to $\omega$ Cen, despite a clear overlap in the overall metallicity distribution and in other elements, and also between them. The majority of clusters have an extremely low $\omega$ Cen membership probability. The red line in Figure~\ref{CDF} represents all 
globular cluster members observed by APOGEE DR17. Although the overwhelming majority of cluster members ($>$ 85$\%$) have an extremely low probability of belonging to $\omega$ Cen, we determine that approximately 10$\%$ of the total members of other clusters could have a probability of 0.1 and only 1$\%$ a probability of 0.6 to be part of $\omega$ Cen. This exercise suggests that, to identify genuine $\omega$ Cen debris in the APOGEE sample, we should favor objects with probabilities greater than 0.6.


In the following sections, we analyze these populations in detail. Focusing on the substructure observed in Figure~\ref{fig_log_prob_vs_feh}, we identify mainly five different populations in this APOGEE sample; what is commonly called the thin/thick disk, the metal-weak thick disk \citep[e.g.,][]{Norris1986,Morrison1990,Beers2002,Robin2014,Anguiano2020}, the halo component with a non-rotating average motion \citep[e.g.,][]{Chiba2000,Hayes2018a}, the above mentioned metal-poor ``plume", and the $\omega$ Cen debris candidates close to P $\sim$ 1. 

\subsubsection{Lindbland Diagram and Orbital Actions}

The combination of chemical consistency with stellar kinematics can help to characterize the $\omega$ Cen tidal debris candidates together with the other sub-structures identified in this exercise. The classical Lindblad diagram ($E$ - $L_{z}$) helps to identify circular orbits from the regions where the orbits are eccentric \citep[e.g.,][]{Sellwood2002}. To further investigate the possible association of $\omega$ Cen debris candidates with the putative cluster, we explored orbital actions ($J_{R}$,$J_{\phi}$,$J_{z}$). 
In axisymmetric potentials, they are integrals of motion and quantify the amount of oscillation of the star along its orbit in the Galactocentric ($R$, $\phi$, $z$) direction, respectively. Furthermore, in an axisymmetric potential, the third action $J_{\phi}$ is the component of the angular momentum around the symmetry axis ($J_\phi$ $\equiv$ $L_{z}$). However, the identification of stellar debris in kinematic-related spaces should be treated with caution. For example, \cite{Jean_Baptiste2017} using high-resolution, $N$-body simulations, showed that caution must be employed before interpreting overdensities in any of those spaces as evidence of relics of accreted satellites. They showed that over-densities of multiple satellites overlap; and satellites of different masses can produce similar substructures. Therefore, reconstructing the $\omega$ Cen accretion event required careful chemical identification of candidates complemented by kinematic information \citep[e.g.,][]{Horta2023}.


We already discussed in Figure~\ref{fig_log_prob_vs_feh} the different substructures we found using a \emph{weak} chemical tagging neural network model together with the \emph{probability} to belong to $\omega$ Cen. We constructed the Lindblad diagram (energy E versus L$_{z}$) to investigate whether there is a kinematic connection between these substructures and the chemically selected $\omega$ Cen debris and the other clusters. The left-hand panel of Figure~\ref{fig_P_limb} investigates the region-dominated by a stellar population within $-$1.9 $<$ [Fe/H] $<$ $-$0.9 and $\omega$ Cen membership probability smaller than 10$\%$. The selected area is colored in orange for all panels. In blue we have the entire sample. The Lindblad diagram (left-hand middle panel) reveals at least three structures. A clean metal-poor population in prograde motion, very likely associated with the metal-weak thick disk \citep[e.g.,][]{Cordoni2021}. Another structure clearly visible in the diagram is a prominent non-rotating high-energetic population. It is widely accepted that this debris is dominated by a merger of a dwarf galaxy as the progenitor, now called ``Gaia-Sausage-Enceladus (GSE)" \citep{Belokurov2018,Haywood2018,Deason2018,Helmi2018,Fattahi2019}. However, using APOGEE \citep{Majewski2017} and GALAH surveys \citep{DeSilva2015}, \cite{Donlon2023} have shown that the stellar populations in the local stellar halo are inconsistent with a scenario in which the inner halo is primarily composed of debris from a single massive, ancient merger event, as has been proposed to explain the GSE structure. 

Finally, there is another non-rotating structure with low energy. \cite{Horta2021} claimed this population as a major building block of the halo buried in the inner Galaxy and called it \emph{Heracles}. Nevertheless, it is relevant to point out that the Milky Way bulge also populates this area of the Lindblad diagram. Furthermore, the existence of bulge metal-poor stars has been reported \citep[e.g.,][and references therein]{Howes2016,Wylie2021}. The right-middle panel in Figure~\ref{fig_P_limb} reveals the radial action $J_{\rm R}$ as a function of the orbital energy ($E$). The three components described above are also visible in this energy-action space; the metal-weak thick disk and the non-rotating low orbital energy population; both show a low$J_{\rm R}$. However, the accreted GSE structure has a large range and very high-$J_{\rm R}$ values compared to the other two main structures. \cite{Belokurov2018} showed that this debris has a strongly radial orbit that merged with the Milky Way at a redshift $z$ $\leq$ 3. All of these structures have a very low probability of being chemically associated to the core of $\omega$ Cen. The lower panel in Figure~\ref{fig_P_limb} shows the [Fe/H] - [Mg/Fe] plane, and illustrate that the vast majority of disk stars are not part of the structures discussed above following the selection using the $\omega$ Cen membership probability and [Fe/H] derived from APOGEE spectra.

We now study the metal-poor ``plume" on the right-hand panels of Figure~\ref{fig_P_limb}. We select this population where $\omega$ Cen membership probability is smaller than 10$\%$ and where $-$2.5 $<$ [Fe/H] $<$ $-$1.9 (see also Fig.~\ref{fig_log_prob_vs_feh}). The energy-action spaces in the middle panels suggest that this is a debris in the Milky Way halo with a very likely accreted origin. This debris, with both prograde and retrograde members and covering areas in these diagrams similar to GSE debris, is chemically independent of GSE from our weak chemical tagging exercise. \cite{Naidu2020} also reported a metal-poor high-$\alpha$ debris, selecting only a retrograde stellar component (l'itoi). Our findings suggest that the progenitor of this debris should be larger than previously reported, showing retrograde and also prograde motion, and probably an important building block of the MW halo.  

\begin{figure*}[ht]
\begin{center}
\includegraphics[width=.48\hsize,angle=0]{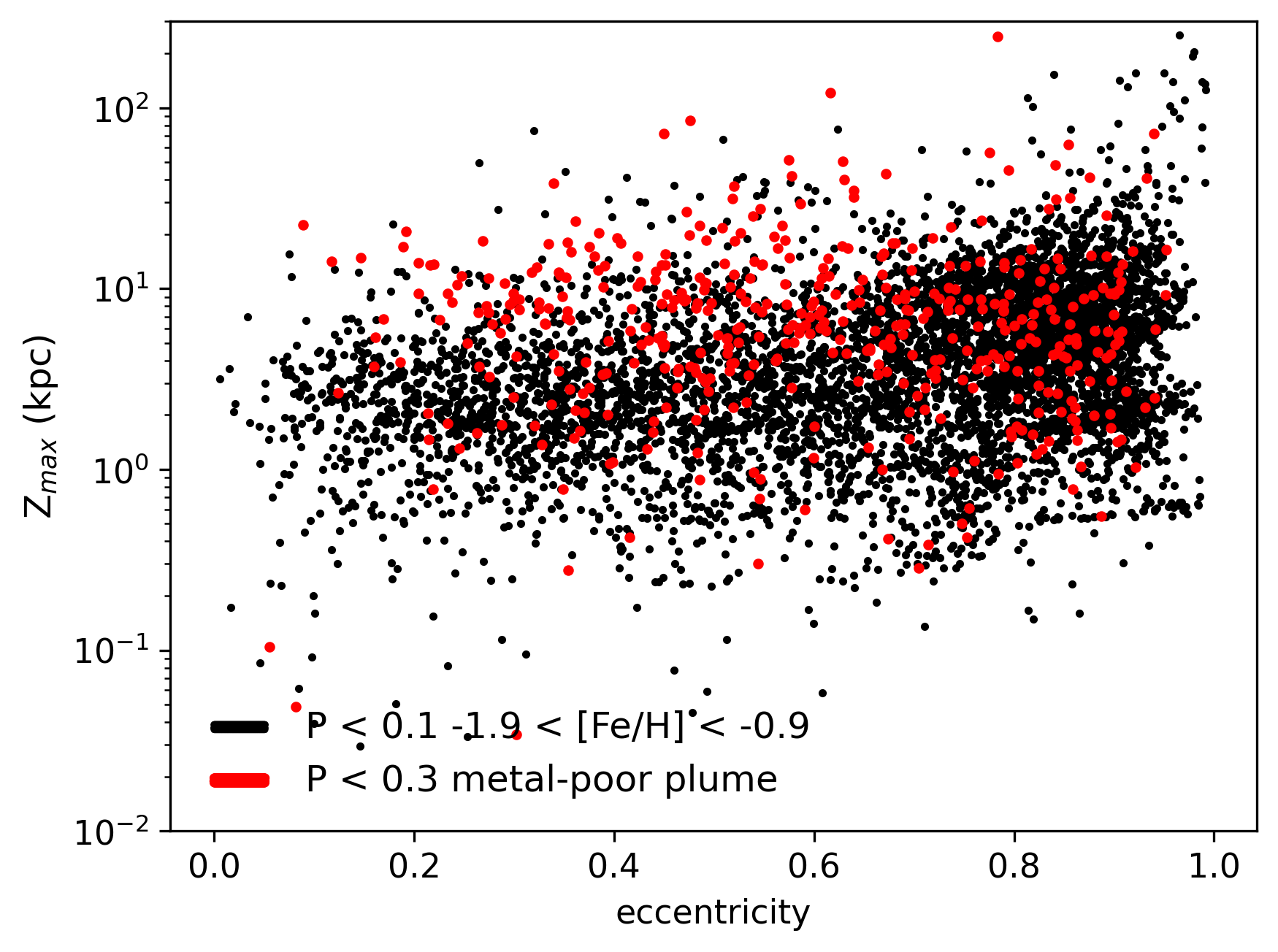}
\includegraphics[width=.48\hsize,angle=0]{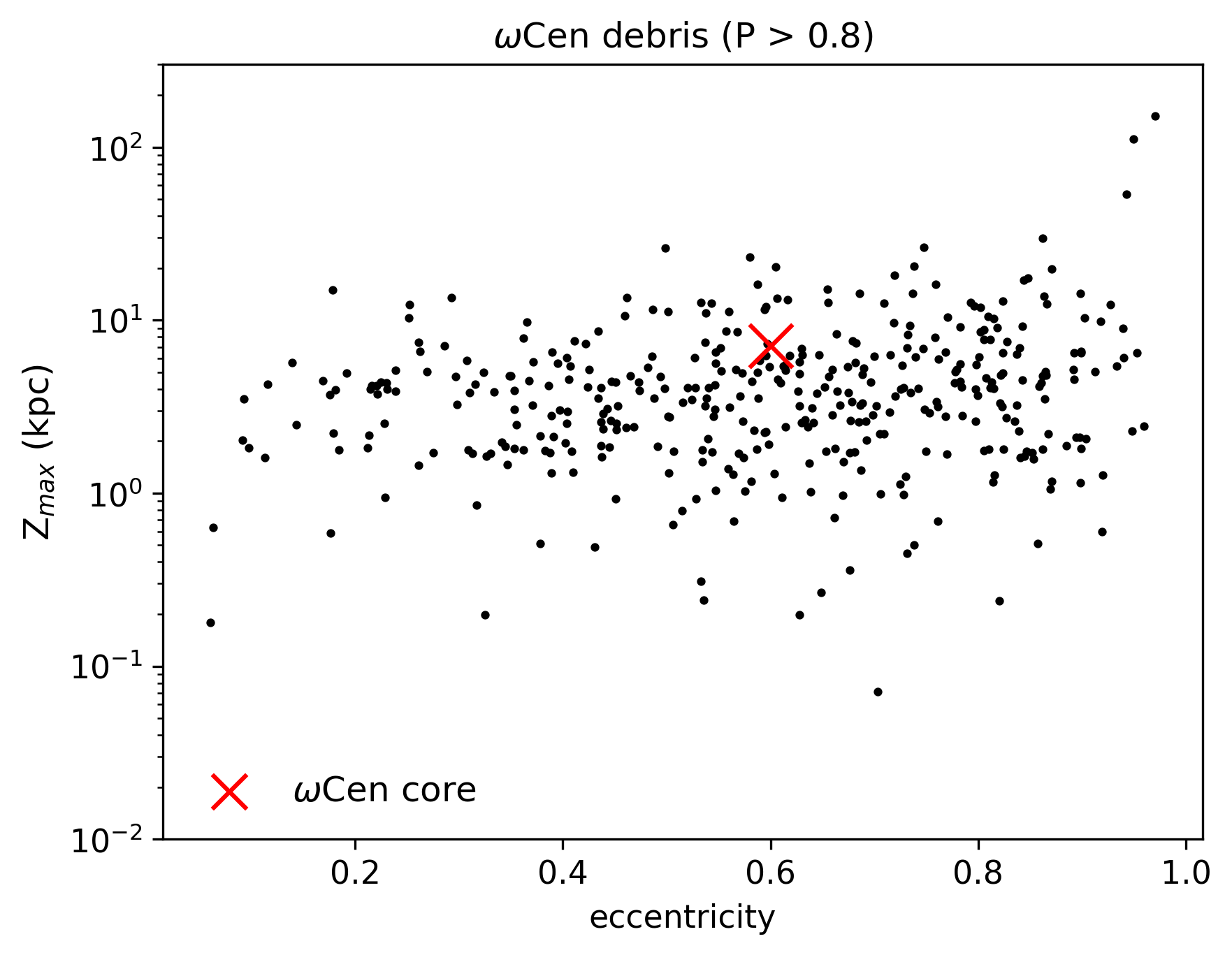}
\includegraphics[width=.48\hsize,angle=0]{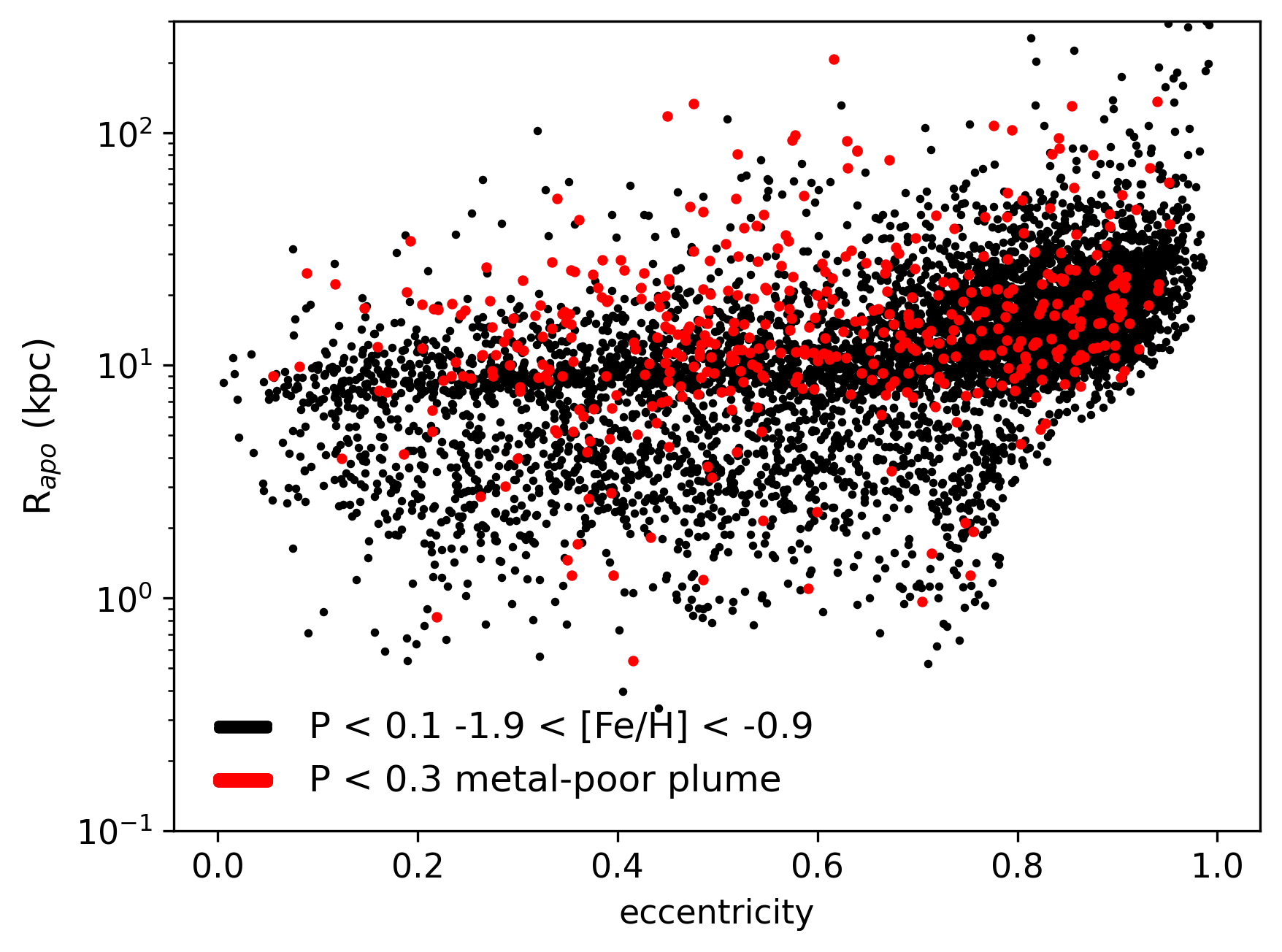}
\includegraphics[width=.48\hsize,angle=0]{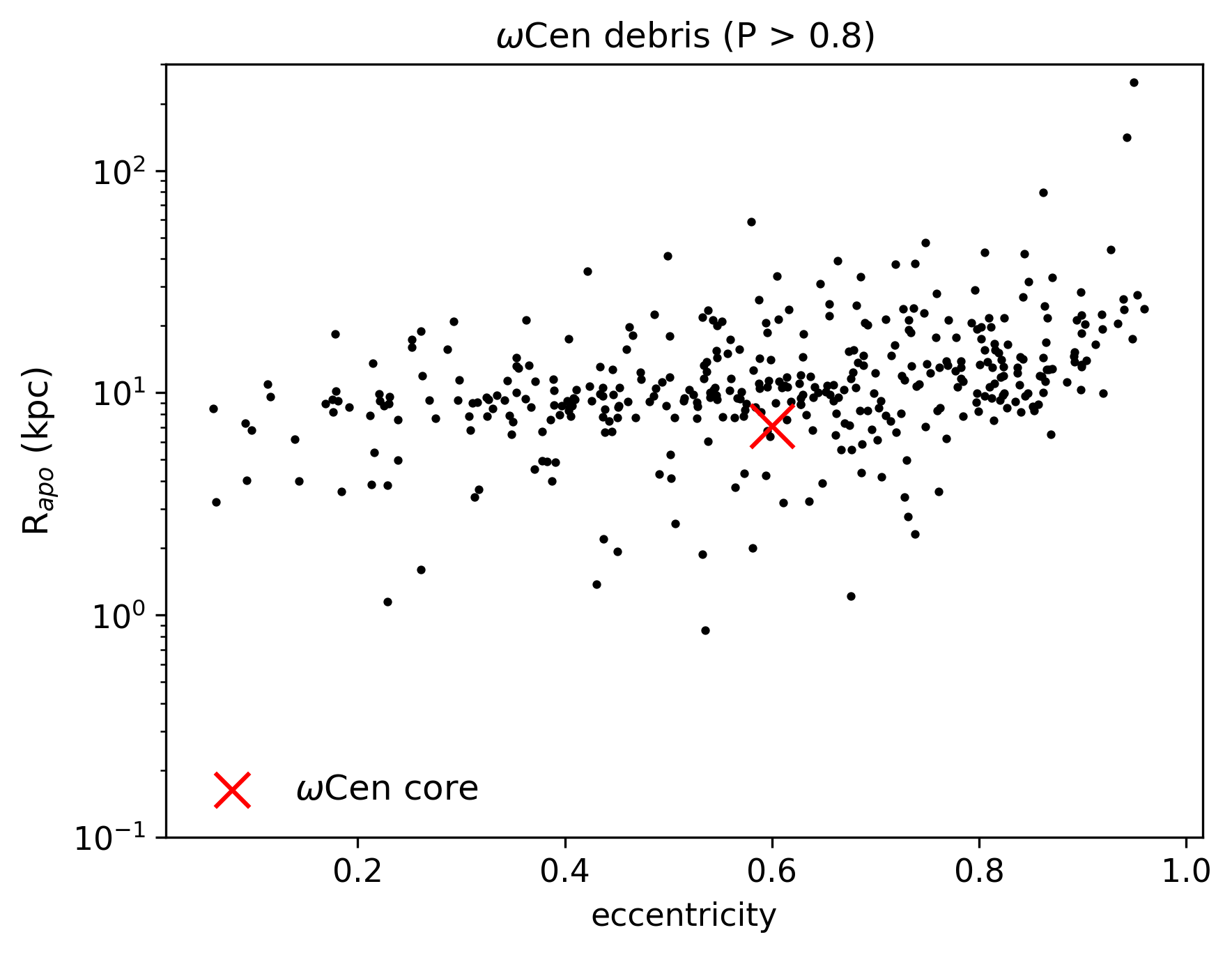}
\end{center}
\caption{In the top panels we have the maximum height achieved above the Galactic plane Z$_{max}$ versus the orbital eccentricities for $P < 0.1$ and -1.9 $<$ [Fe/H] $<$ -0.9 (left) and $P > 0.8$ (right), where $\omega$ Cen core is marked with a red X. The lower panel also shows the apocenter with respect to the eccentricity. See text for details.}
\label{fig_eccen}
\end{figure*}

We finally focus on the $\omega$ Cen debris candidates. Figure~\ref{fig_P_ocen} shows the energy-action spaces for a population with $P$ $>$ 0.8. Following the Lindblad diagram, some of the debris have prograde and retrograde disk-like kinematics, whereas most of the debris lie within the accreted halo, showing high-$J_{\rm R}$ values. We notice that GSE, the metal-poor ``plume" and the $\omega$ Cen debris candidates show a very similar energy-action spaces behavior. 
While they are all part of the accreted halo, they are chemically distinct from our chemical tagging exercise, suggesting that they are independent events that build up the Milky Way stellar halo. However, $\omega$ Cen exhibits evidence of chemical enrichment and could be the inner-most part of the progenitor system. Hence, the chemical tagging performed in this study cannot definitely say that GSE and $\omega$ Cen have distinct origins. In the next section, we explore a few orbital parameters of the $\omega$ Cen debris candidates in more detail. 


\begin{figure}[ht]
\begin{center}
\includegraphics[width=1.\hsize,angle=0]{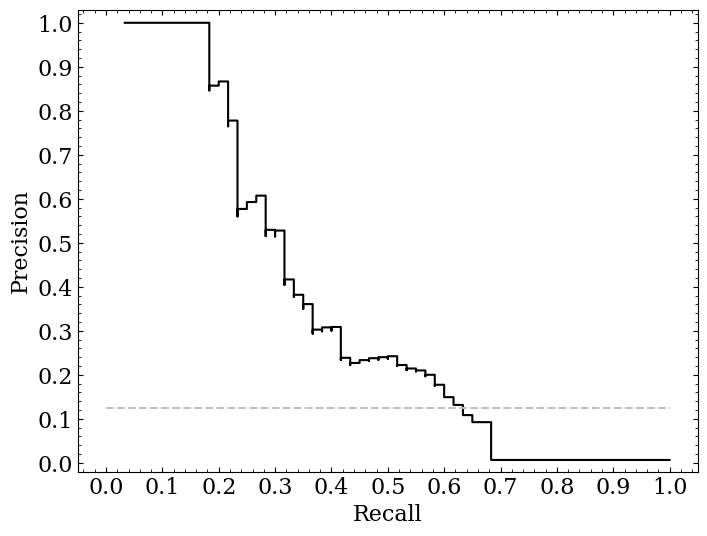}
\end{center}
\caption{Best model recall and precision are plotted for P $>$ P$_{thresh}$ from 0 to 1 at intervals of 0.01. The dashed grey horizontal line at 0.12 represents a random uneducated model. The model is not helpful at discriminating M54, therefore any predictions would not be meaningful.}
\label{m54}
\end{figure}

\subsubsection{Orbital Parameters}

Using the selected populations described in the previous section, where we used the $\omega$ Cen membership probability and [Fe/H], we investigate the stellar eccentricity, vertical excursion from the Galactic plane ($Z_{max}$) and apocenter ($R_{apo}$).  

The top panels in Figure~\ref{fig_eccen} show the maximum height achieved above the Galactic plane Z$_{max}$ versus the orbital eccentricities for $P < 0.1$ (left) and $P > 0.8$ (right), respectively. For the stellar population where $-$1.9 $<$ [Fe/H] $<$ $-$0.9 and $\omega$ Cen membership probability is smaller than 10$\%$, we find a group with high orbital eccentricity ($e > 0.8$), that also tends to have larger vertical excursions. These populations are associated with GSE. The low orbital eccentricity population is a mixture of different origins, possible remnants of accretion events, the metal-weak thick disk, and the MW bulge \citep[e.g.;][]{Mackereth2019,Queiroz2021} There is also an apocenter pile-up (see bottom left panel in Figure~\ref{fig_eccen}), in both the sample of stars with high and low orbital eccentricities. \cite{Deason2018} suggested a massive dwarf
progenitor, especially for the high eccentricity sample. 

\begin{figure*}[ht]
\begin{center}
\includegraphics[width=.93\hsize,angle=0]{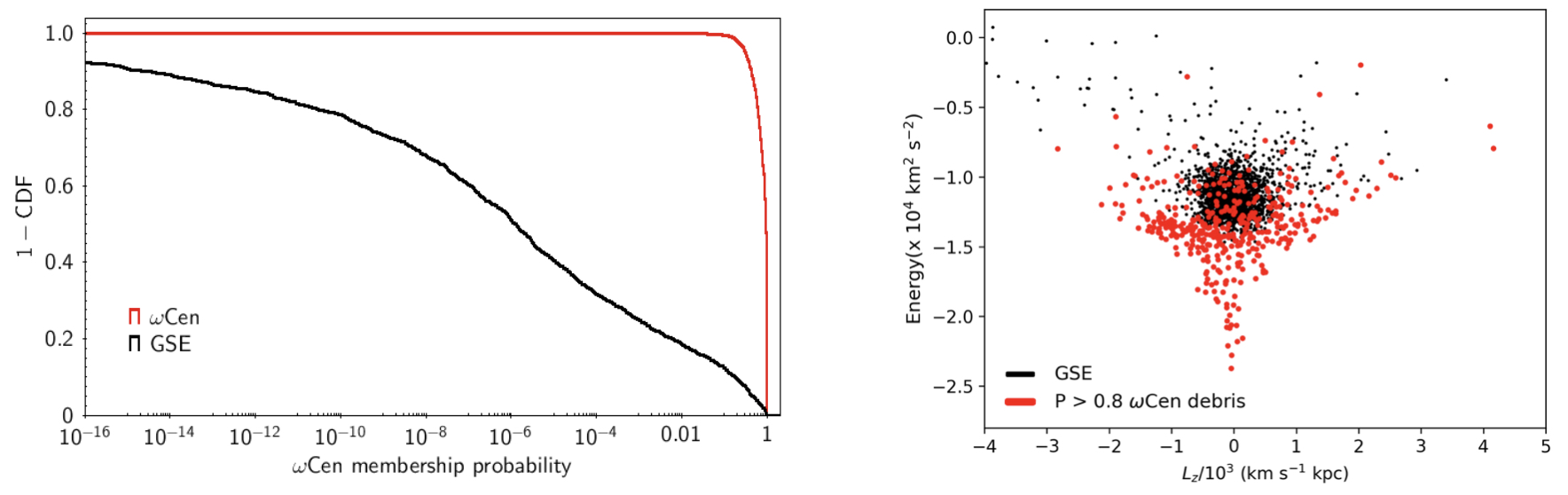}
\end{center}
\caption{\emph{Left panel:} Reverse cumulative distribution functions for the $\omega$ Cen membership probability for the selected GSE sample (blue line) and the $\omega$ Cen members (green line).  Note the ``bump" in the GSE sample for $P > 0.001$, those are $\omega$ Cen debris members and the metal-poor plume that made it into the GSE selection. More than 90$\%$ of the GSE members have a $P < 0.1$ to have a similar chemical pattern than $\omega$ Cen cluster. \emph{Right panel:} Lindblad diagram for the selected GSE sample (black dots) and the $\omega$ Cen debris candidates (red dots). The kinematical overlap between the different structures is not evident. Furthermore, the selected GSE sample do not show stars at low energy.}
\label{fig_GSE_ocen}
\end{figure*}

The star associated with $\omega$ Cen debris for a P $>$ 0.8 show a wide range of Z$_{max}$ (0.5 $<$ Z$_{max}$ $<$ 100 kpc), together with a large range in eccentricity (0.1 $<$ $e$ $<$ 1.0). In Figure~\ref{fig_eccen} we also have the position of $\omega$ Cen core marked as a red cross ($e$ $\sim$ 0.6, Z$_{max}$ $\sim$ 2 kpc). 

\section{Is $\omega$ Centauri the stellar nucleus of the Gaia-Sausage-Enceladus (GSE) galaxy?}

\cite{Myeong2018} reported the properties of eight GCs, including $\omega$ Cen, that are consistent with being associated with the merger event that gave rise to the GSE. Furthermore, \cite{Massari2019,Pfeffer2021} further suggested that $\omega$ Cen could be the surviving nuclear star cluster of GSE. \cite{Forbes2020} assigned NGC 1851 as the nuclear star cluster of GSE, and $\omega$ Cen with Sequoia \citep{Myeong2019}. \cite{Limberg2022} estimated the stellar mass of the progenitor of $\omega$ Cen to be M $\approx$ 1.3 $\times$ 10$^{9}$ M$_{\odot}$, well within literature expectations for GSE. This leads the authors to envision GSE as the best available candidate for the original host galaxy of $\omega$ Cen. Note that, \cite{Donlon2023} reported that the structure called the GSE is not really a single ancient massive merger, they found that it is instead built up from a combination of many smaller merger events.

In this section, we use the $\omega$ Cen debris candidates identified in this study to further understand its progenitor and its possible association with the structure reported in the Milky Way halo. 

\subsection{Building a GSE Candidates Sample}

Creating a pure GSE sample that is free from contamination by other structures is challenging. \cite{Carrillo2023} explored selections made in eccentricity, energy-angular momentum (E-L$_{z}$), radial action-angular momentum (J$_{r}$-L$_{z}$), action diamond, and [Mg/Mn]-[Al/Fe] in the observations, see also \cite{Horta2023}. In all these selections, they reported contamination from in-situ and other accreted stars. 

For the purpose of this study, we use the parent sample of the GSE defined in \cite{Horta2023}, based on the (E-L$_{z}$) plane. Note that a good number of $\omega$ Cen debris and the metal-poor ``plume" are also within this selection. We stress that there is no guarantee that this selection of GSE provides an unbiased sample of the stars of the accreted galaxy \citep[e.g.;][]{Feuillet2020,Bonifacio2021}.

Figure~\ref{fig_GSE_ocen} shows the reverse cumulative distribution function of the $\omega$ Cen membership probability for the selected GSE members and $\omega$ Cen cluster, and on the right panel, the (E-L$_{z}$) plane for $\omega$ Cen debris candidates with P $>$ 0.8 (red points) and the parent sample of the GSE (black points). The potential contamination from the metal-poor ``plume" and the $\omega$ Cen is evident in the GSE selected sample for P $>$ 0.001. Almost all GSE stars in our sample have a $\omega$ Cen membership probability smaller than 0.1. Our results indicate that the selected GSE members are chemically distinct from the current $\omega$ Cen core, and hence the $\omega$ Cen debris identified in this study. We would expect a similar chemical evolution enrichment history for a large number of GSE members, as the progenitor galaxy and the putative cluster and its debris. However, it is well established that $\omega$ Cen shows clear signs of its own self-enrichment, e.g., GC-like chemical evolution producing large anti-correlations \citep{Carretta2019,Garay2022}. In fact, \cite{Romano2007} showed that in a closed-box self-enrichment scenario, the MDF of the clusters cannot be reproduced. However, the main chemical properties of $\omega$ Cen are matched if it is the compact remnant of a dwarf spheroidal galaxy that evolved in isolation and then accreted by the MW. \cite{Romano2007} assumed a relatively long-lasting star formation activity (though with most of the stars forming within 1 Gyr), standard IMF and standard stellar yields, their models satisfactorily reproduce several observed abundance ratios as a function of [Fe/H], proving that the ingested satellite would also have similar chemical properties to $\omega$ Cen. To make this scenario even more intricate, the GSE sample could be the product of many smaller mergers with a wide range of dynamical times \citep[e.g.,][]{Donlon2023, Donlon2024}.

\subsection{M54 + Sgr = $\omega$ Cen?}

Hubble Space Telescope photometry of the massive M54 cluster shows multiple turnoffs coming from the cluster population and Sgr, indicating different star formation epochs \citep{Siegel2007}. In this regard, \cite{Carretta2010} found that by including stars of the Sgr nucleus in their M54 spectroscopic sample, the metallicity distribution is similar to that observed in $\omega$ Cen. \cite{Carretta2010} concluded that M54 and $\omega$ Cen are nuclear clusters in dwarf galaxies showing a different phase of their dynamical evolution.

Following this idea, we performed a similar training and probability estimation as described for $\omega$ Cen but now trained in M54 APOGEE members \citep{schiavon2024} and left the APOGEE Sgr stream objects \citep{Hayes2020} out of the negative sample. Our goal is to see what fraction of Sgr stars are chemically tagged to M54. As we described in Section~\ref{sec:chemtagmodel}, Figure~\ref{m54} shows the performance of the model in terms of recall (the proportion of cluster stars that were correctly identified as such - the true positive rate) and precision (the proportion of positive labels that correctly marked positive). Figure~\ref{m54} indicates that the model is not good at differentiating M54, therefore any predictions would not be meaningful.

There are a large number of pieces of evidence that relate the M54 cluster to Sgr dSph \citep[e.g.,][]{SL2000,Siegel2007,Bellazzini2008}. However, in this study, we found that our neural network modeling (see Sect.~\ref{sec:chemtagmodel}) is not able to chemically associate the core of M54 with those objects that are part of the Sgr stream, suggesting that most of the stars in the progenitor galaxy do not show the same chemical properties as M54. However, M54 APOGEE sample is potentially contaminated from the main body of the Sgr dSph. Additionally, \cite{Hayes2020} demonstrated the presence of a chemical composition gradient in the Sgr dSph, suggesting that even the galaxy's main body may not align with its streams. 

\subsection{Orbital Parameters}

In this section, we study the kinematical structure of the parent sample of the candidates for the GSE \citep{Horta2023} and the $\omega$ Cen debris candidates using the Lindblad diagram. $N$-body simulations show that a massive disk galaxy merging with the MW can generate debris with a complex phase-space structure, a large range of orbital properties, and a complex range of chemical abundances \citep{Villalobos2009,Jean_Baptiste2017,Koppelman2020}. 

In the right panel of Figure~\ref{fig_GSE_ocen} we have the resulting angular momentum - orbital energy distribution for the GSE sample (black dots) and \emph{$\omega$ Cen} debris candidates (red dots). Most of the $\omega$ Cen debris tend to pile up in a narrow range of Energy with a large angular momentum range. We also have debris candidates with low energy values and non-rotating kinematics. A comparison with the GSE sample need to be made carefully, as this sample was selected using the (E-L$_{z}$) plane, hence biasing kinematically the sample. Figure~\ref{fig_GSE_ocen} shows that the high energy \emph{$\omega$ Cen} debris share similar angular momentum values with the GSE sample. However, our results indicate that there is no clear chemical and kinematical overlap between the \emph{$\omega$ Cen} debris and the GSE sample.


\section{Numerical simulations predictions}


In this section, we study the predictions of the tidal disruption of the $\omega$ Cen progenitor using an $N$-body simulation. In particular, we discuss the global distribution of the components stripped by the tidal effects, and we compare the position and kinematics of the $\omega$ Cen debris identify in this study with the orbital motion of the progenitor derived in the simulations. $\omega$ Cen has singular characteristic, and there is evidence that $\omega$ Cen may be the remnant of a heavily stripped dwarf galaxy, as described in detail in Section~\ref{sec:intro}. A detailed modeling of its progenitor remains challenging, mainly due to the difficulty of simulating the long-term orbital evolution and being able to mimic the current orbital elements of the cluster, as the evolution of the Galactic gravitational potential and the effect of dynamical friction are not negligible \citep[e.g.,][]{Meza2005}

\subsection{Methodology}

In this study, we focus on the global distribution of the tidally stripped stars using a simple progenitor model. We first integrated the orbit backward in time in the MW gravitational potential from the current position of $\omega$ Cen using a test particle. We recovered the position and velocity of the cluster 500~Myr ago. The apocentric and pericentric distances are 7.1~kpc and 1.6~kpc, respectively, and it has experienced five pericentric passages over the last 500~Myr. During this time, we have assumed that the Galactic potential does not change with time. We then simulate the tidal disruption of a progenitor cluster using an $N$-body simulation. We construct the progenitor dwarf galaxy as a Plummer model using \textsc{MAGI} \citep{Miki2018}. The total mass and scale radius of the Plummer distribution are set to $10^8 M_{\odot}$ and 1 kpc, respectively. We assume that the dark matter halo of the progenitor galaxy has already been stripped at the beginning of the simulation. The total number of particles is 1,048,576, and we use the gravitational octree code \textsc{GOTHIC} \citep{Miki2017} to run the simulation using the accuracy parameter of $\theta=0.5$. We adopt the Plummer softening parameter of 16 pc. 



\subsection{Results}

\begin{figure}[ht]
\begin{center}
\includegraphics[width=1.\hsize,angle=0]{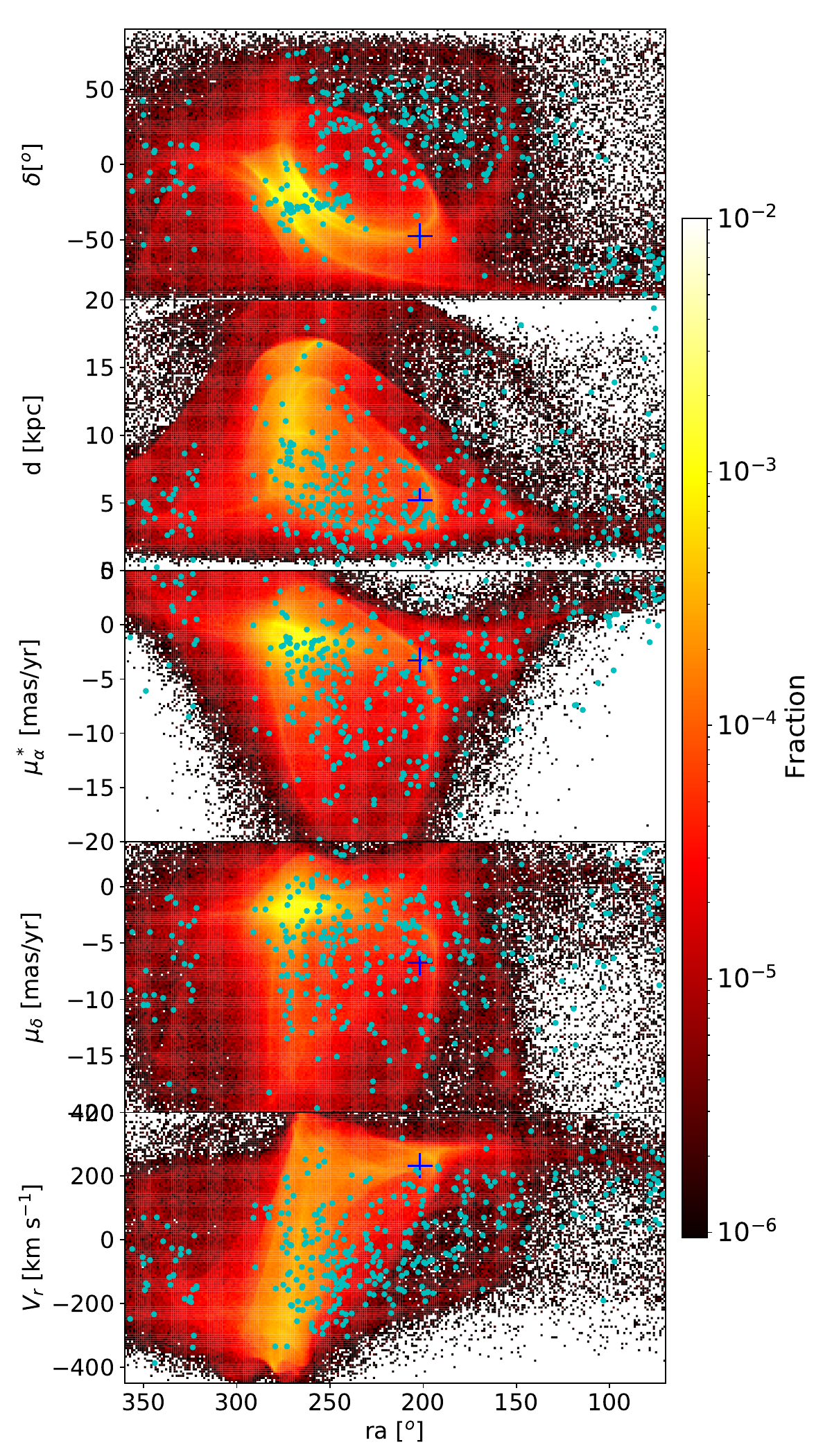}
\end{center}
\caption{Distribution of the simulated $\omega$ Cen debris in (a) declination, (b) heliocentric distance, proper motion in (c) $\alpha$ and (d) $\delta$,  and (e) radial velocity as a function of $\alpha$. The colors represent the mass fraction in each region normalized by the total mass of the $\omega$ Cen progenitor. The cyan points indicate the $\omega$ Cen membership candidates where the $P > 0.8$. The blue cross indicates the observed information of the $\omega$ Cen cluster. 
}
\label{fig_simresults}
\end{figure}

\begin{figure}[ht]
\begin{center}
\includegraphics[width=1.\hsize,angle=0]{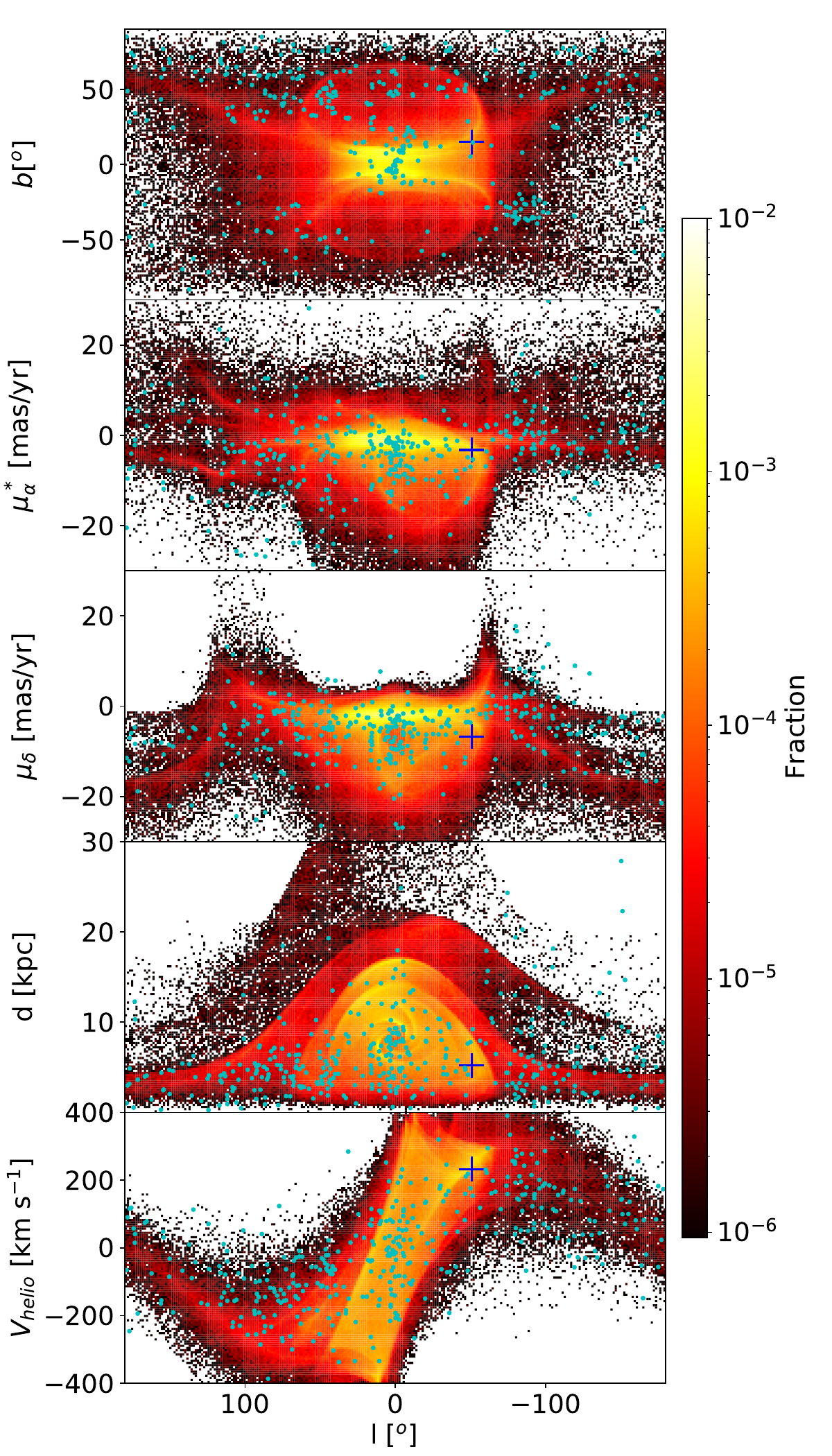}
\end{center}
\caption{Distribution of the simulated $\omega$ Cen debris in (a) Galactic latitude, proper motion in (b) $\alpha$ and (c) $\delta$, (d) heliocentric distance, and (e) heliocentric velocity as a function of Galactic longtitude. The colors represent the mass fraction in each region normalized by the total mass of the $\omega$ Cen progenitor. The cyan points indicate the $\omega$ Cen membership candidates where the $P > 0.8$. The blue cross indicates the observed information of the $\omega$ Cen cluster. 
}
\label{fig_simresults2}
\end{figure}

Figure~\ref{fig_simresults} and Figure~\ref{fig_simresults2} show the simulated distribution of $\omega$ Cen debris. The overplotted cyan points indicate the $\omega$ Cen membership candidates where $P > 0.8$. The tidally stripped $\omega$ Cen debris obtained from the simulation covers the spatial distribution of the observed $\omega$ Cen candidates. In addition, our simulation nicely reproduces the extent of the distribution of the candidates in terms of distance, proper motion, and radial velocity. The number density of debris is high around (l,b)=$(270^{\circ}, -30^{\circ})$, that area is where APOGEE observed the TESS CVZ \citep{Beaton2021,Santana2021}. Multiple $\omega$ Cen candidates are present along high-density streams. Beyond 10~kpc, the APOGEE data are poor. Therefore, it is acceptable that there are few $\omega$ Cen candidates in this region, even though we can see several structures derived from $\omega$ Cen in the simulation. The number density of $\omega$ Cen debris is high around ($ra$, d)=(260$^{\circ}$, 15 kpc), which could be a target for future observations of $\omega$ Cen tidal debris. Multiple streams around $\omega$ Cen, indicated by the blue plus, appear in the simulation. However, it should be noted that the APOGEE data around the $\omega$ Cen are not sufficient. In particular, the regions $120^{\circ}<$ra$<250^{\circ}$ and $\delta< -10^{\circ}$ have few APOGEE data. In addition, the streams have a large spread in proper motions, making it challenging to find the structure simply from the Gaia observations. For the regions of $100^{\circ}<$ra$<250^{\circ}$ and $\delta> -10^{\circ}$, the mass fraction of debris in the simulation is relatively small, while the $\omega$ Cen candidates are numerous. This is likely due to the richness of the APOGEE data in this region and the relatively close distances to the debris. It suggests that a large amount of $\omega$ Cen debris is still hidden.

\begin{figure}[ht]
\begin{center}
\includegraphics[width=1.\hsize,angle=0]{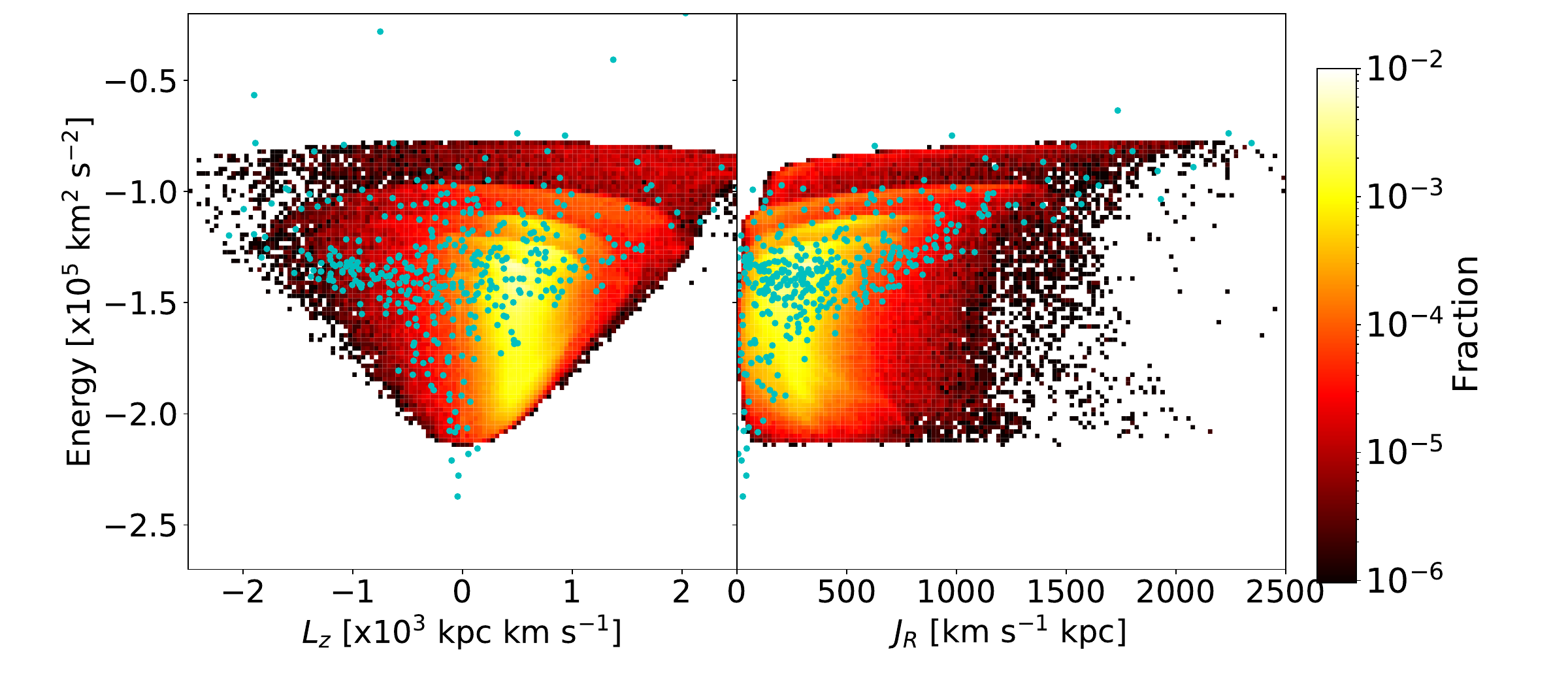}
\end{center}
\caption{Left panel: the distribution of the $\omega$ Cen debris in the Lindblad diagram. Right panel: the distribution in the energy-action space. 
The cyan points indicate the $\omega$ Cen membership candidates where the P $>$ 0.8. 
}
\label{fig_simresults4}
\end{figure}

The left panel of Figure~\ref{fig_simresults4} shows the Lindblad diagram for the simulated $\omega$ Cen debris, and in this figure we can also see that the tidal debris is distributed over a wide range in $L_z$. The major population has a retrograde orbit, but some debris has prograde kinematics. Although the integration time is slightly short, we also find an over-density in the E-$L_z$ plane. This over-density lies in a similar or slightly lower energy region as the GSE (see right panel in Figure~\ref{fig_GSE_ocen}). It should be noted that the relics of multiple accreted satellites may appear in similar positions in the E-$L_z$ plane, as pointed out by \citet{Jean_Baptiste2017}. Hence, our approach with chemical tagging is powerful in approaching the origin of the substructures. The right panel of Figure~\ref{fig_simresults4} shows the radial action as a function of orbital energy. In addition to the E-$L_z$ distribution, the simulated distribution of tidal debris in the E-$J_R$ plane matches the scatter on the observed data. The $\omega$ Cen debris shows high $J_R$ values, indicating that the debris has strong radial orbits. 

According to $\omega$ Cen debris candidates obtained using the characteristics of the chemical abundance patterns, we find a non-negligible prograde-orbiting stars (see Figure~\ref{fig_P_ocen}). The wide distribution on the $L_z$-axis requires that when the progenitor passes through the pericenter, a part of the progenitor has to pass on the opposite side of the Galactic center to the main component. For our adopted gravitational potential \citep{Bovy2015}, the pericentric distance is about 1~kpc; therefore, a progenitor model with a radius of more than 1~kpc is required to obtain a broad distribution on the $L_z$-axis. Concerning the ratio of prograde to retrograde components, this suggests a strong dependence on the size of the progenitor and its density profile. In fact, a simulation with the current $\omega$ Cen-like progenitor (with a stellar mass of $10^6 M_{\odot}$ and a core radii of 76 pc) produces a distribution concentrated in the $L_z>0$ region. Indeed, we cannot rule out the possibility that there have been periods of extremely small pericentric distances as a result of the time evolution of the Galactic potential. However, it should be emphasized that it becomes more difficult for the $\omega$ Cen core to survive in such cases. That is, the presence of a prograde component suggests that the progenitor was a dwarf galaxy with a somewhat large stellar mass ($\gtrsim 10^8 M_{\odot}$).

\section{Summary \& Conclusions} \label{sec:conclusions}

In this study we use data from the APOGEE survey \cite{Majewski2017} included in the SDSS Data Release 17 (DR17, \citealt{SDSS_DR17}). We also exploit \emph{Gaia} DR3 \citep{Gaia2022}, and the APOGEE StarHorse value-added catalog \citep{SH2020}. A total of 1,872 stars in the core of $\omega$ Cen have APOGEE spectra \citep{Santana2021,Meszaros2021}. Having the stellar individual abundances of the $\omega$ Cen core as a training sample and using chemical tagging from a neural network modeling we identify $\omega$ Cen debris candidates in the APOGEE survey. We test our neural network modelling using four different clusters observed with APOGEE containing more than 200 individual objects, those are NGC 104, NGC 3101, NGC 2808 and NGC 6121. Despite the clear overlap in metallicity and in other elements between $\omega$ Cen and these clusters, the model is able to show how chemically unique is every cluster with respect to $\omega$ Cen. Stellar members of these clusters have a very low $\omega$ Cen membership probability. We found 463 $\omega$ Cen debris candidates with a probability P $>$ 0.8. Remarkably, we reported that a good number of them show C + N, Al and Ce abundances enhanced with respect to other debris candidates and most Galactic halo stars. 

A chemo-dynamical study of $\omega$ Cen debris candidates shows that most of them lie within the accreted MW halo, however, debris with prograde and retrogade disk-like kinematics are also found. Most of the debris also have high radial action values. \cite{Massari2019, Pfeffer2021} suggested that $\omega$ Cen was the nuclear star cluster of GSE. To further investigate this potential association we have compared the APOGEE parent sample of the GSE defined in \cite{Horta2023} with $\omega$ Cen. We found that nearly all the GSE stars in our sample show a $\omega$ Cen membership probability smaller than 0.1. This result indicates that the actual $\omega$ Cen core and the selected GSE members are chemically distinct. However, $\omega$ Cen shows clear evidences of its own self-enrichment, hence, the stars in its progenitor galaxy should have a different chemical evolution than the one we observe today in the core of $\omega$ Cen. 

Despite the negative result associating GSE and $\omega$ Cen via chemical tagging modeling, we cannot rule out their potential association. The kinematical exercise findings are inconclusive as well, however, most of the $\omega$ Cen debris found in this study using stellar individual abundances lie in a different area covered by the GSE sample. There is some overlap with the GSE sample, but this is likely due to contamination from the $\omega$ Cen debris, as avoiding the GSE's influence when selecting any halo sample is nearly impossible. We also found a metal-poor ``plume" with [Fe/H] $<$ -2.0 in the probability - metallicity plane (see Sect.~\ref{debris_candi}). This population has low probability of following the chemical pattern of $\omega$ Cen but it is distinct from the other structures like GSE. The energy-actions diagram shows that this population is very likely accreted and probably part of an important building block of the MW halo.

Finally, we have compared our observational results with the tidal disruption of the $\omega$ Cen progenitor using N-body simulations. We found that the properties of the debris from the simulations span the spatial distribution of the observed $\omega$ Cen candidates. Furthermore, the N-body simulations results suggested progenitor was a dwarf galaxy with a somewhat large stellar mass ($\gtrsim 10^8 M_{\odot}$).

\section*{Acknowledgements}

BA acknowledges support by the Spanish Ministry of Science, Innovation and Universities and the State Research Agency (MICIU/AEI) with the grant RYC2022-037011-I and by the European Social Fund Plus (FSE+). T.K. was supported in part by JSPS KAKENHI Grant Number JP22K14076. Numerical simulations were performed in part using Cygnus at the CCS, University of Tsukuba. YH was supported in part by JSPS KAKENHI Grant Numbers JP22KJ0157, JP25H00664, JP25K01046, MEXT as ``Program for Promoting Researches on the Supercomputer Fugaku" (Structure and Evolution of the Universe Unraveled by Fusion of Simulation and AI; Grant Number JPMXP1020230406), and JICFuS. CAP acknowledges support from the Spanish government through grants AYA2017-86389-P and PID2020-117493GB-100. This work has made use of data from the European Space Agency (ESA) mission {\it Gaia} (\url{https://www.cosmos.esa.int/gaia}), processed by the {\it Gaia} Data Processing and Analysis Consortium (DPAC, \url{https://www.cosmos.esa.int/web/gaia/dpac/consortium}). Funding for the DPAC has been provided by national institutions, in particular the institutions participating in the {\it Gaia} Multilateral Agreement. This research has made use of NASA's Astrophysics Data System Bibliographic Services. Funding for the Sloan Digital Sky Survey IV has been provided by the Alfred P. Sloan Foundation, the U.S. Department of Energy Office of Science, and the Participating Institutions. SDSS-IV acknowledges support and resources from the Center for High-Performance Computing at the University of Utah. The SDSS web site is (\url{https://www.sdss.org}). SDSS is managed by the Astrophysical Research Consortium for the Participating Institutions of the SDSS Collaboration including the Brazilian Participation Group, the Carnegie Institution for Science, Carnegie Mellon University, the Chilean Participation Group, the French Participation Group, Harvard-Smithsonian Center for Astrophysics, Instituto de Astrof\'isica de Canarias, The Johns Hopkins University, Kavli Institute for the Physics and Mathematics of the Universe (IPMU) / University of Tokyo, Lawrence Berkeley National Laboratory, Leibniz Institut f\"ur Astrophysik Potsdam (AIP), Max-Planck-Institut f\"ur Astronomie (MPIA Heidelberg), Max-Planck-Institut f\"ur Astrophysik (MPA Garching), Max-Planck-Institut f\"ur Extraterrestrische Physik (MPE), National Astronomical Observatories of China, New Mexico State University, New York University, University of Notre Dame, Observat\'orio Nacional / MCTI, The Ohio State University, Pennsylvania State University, Shanghai Astronomical Observatory, United Kingdom Participation Group, Universidad Nacional Aut\'onoma de M\'exico, University of Arizona, University of Colorado Boulder, University of Oxford, University of Portsmouth, University of Utah, University of Virginia, University of Washington, University of Wisconsin, Vanderbilt University, and Yale University.

%


\bibliographystyle{aasjournals}
\bibliography{ref_og.bib}



\end{document}